\documentclass[preprint,12pt,authoryear]{elsarticle}

\usepackage{amssymb}

\usepackage[linesnumbered,vlined,
ruled,titlenumbered 
]{algorithm2e}
\usepackage{changepage}
\usepackage{amssymb}
\usepackage[figuresright]{rotating}
\usepackage{subfigure} 
\usepackage{epsfig}    
\usepackage{graphicx,psfrag,epsf}
\usepackage{epstopdf}
\usepackage{url}
\usepackage{color}
\usepackage{multirow}
\newtheorem{theorem}{Theorem}
{Lemma}
{Proposition}
\newtheorem{corollary}
{Corollary}
{Example}

\begin{document}
\def\letas{\mathrel{\mathop{=}\limits^{\triangle}}}
\def\ind{\begin{picture}(9,8)
         \put(0,0){\line(1,0){9}}
         \put(3,0){\line(0,1){8}}
         \put(6,0){\line(0,1){8}}
         \end{picture}
        }
\def\nind{\begin{picture}(9,8)
         \put(0,0){\line(1,0){9}}
         \put(3,0){\line(0,1){8}}
         \put(6,0){\line(0,1){8}}
         \put(1,0){{\it /}}
         \end{picture}
    }
\def\Box{\begin{picture}(9,8)
        \put(0,0){\line(1,0){8}}
        \put(0,8){\line(1,0){8}}
        \put(0,0){\line(0,1){8}}
        \put(8,0){\line(0,1){8}}
        \end{picture}
        }

\newcommand{\trace}{\mathop{\mathrm{tr}}}

\begin{frontmatter}



\title{A generalized EMS algorithm for model selection with incomplete data}


\author[label1]{Ping-Feng Xu\corref{cor1}}\author[label1]{Lai-Xu Shang}
    \author[label2]{Man-Lai Tang}
    \author[label3]{Na Shan}
    \author[label4]{Guoliang Tian}
\address[label1]{School of Mathematics and Statistics, Changchun University of Technology, Changchun, China}
\address[label2]{Department of Mathematics and Statistics, Hang Seng University of Hong Kong, Hong Kong}
\address[label3]{School of Psychology, Northeast Normal University, Changchun, China}
\address[label4]{Department of Mathematics, Southern University of Science and Technology, Shenzhen, China}
\cortext[cor1]{Corresponding author. Email addresses:
\url{xupf900@gmail.com}}

\begin{abstract}
Recently, a so-called E-MS algorithm was developed for model selection in the presence of missing data. Specifically, it performs the Expectation step (E step) and Model Selection step (MS step) alternately to find the minimum point of the observed generalized information criteria (GIC).
In practice, it could be numerically infeasible to perform the MS-step for high dimensional settings.
In this paper, we propose a more simple and feasible generalized EMS (GEMS) algorithm which simply requires a decrease in the observed GIC in the MS-step and includes the original EMS algorithm as a special case. We obtain several numerical convergence results of the GEMS algorithm under mild conditions. We apply the proposed GEMS algorithm to Gaussian graphical model selection and variable selection in generalized linear models and compare it with existing competitors via numerical experiments. We illustrate its application with three real data sets.
\end{abstract}


\end{frontmatter}


\section{Introduction}

Information criteria based on observed log-likelihood function is usually used for model selection and can be readily computed via the famous Expectation-Maximization (EM) algorithm, which was first proposed by \citet{DempsterLR1977} to compute the maximum likelihood estimates in the pesence of incomplete data. For examples, \citet{BuesoQA1999} computed the minimum description length (MDL) using the EM algorithm.
\citet{ClaeskensConsentino2008} proposed several variations based on the
\textsc{aic} for variable selection in the presence of missing covariate.
\citet{IbrahimZT2008} proposed a IC$_{H,Q}$ criteria which depends only on the output from the EM algorithm to compute the observed likelihood.

As pointed out by \citet{JiangNR2015},
the EM approach usually  leads to the notorious ``double-dipping" problem as ones will use the assumed model twice, i.e., once in the measure of lack-of-fit (i.e., the negative log-likelihood) and once in the conditional
expectation of this measure in the EM algorithm.  The multiple usage of the assumed model has been shown in the literature to bring false supporting evidence for an incorrect model (see, \citet{CopasEguchi2005} and \citet{JiangNR2011}). To avoid the aforementioned problem,
\citet{JiangNR2015} generalized the EM algorithm to the EMS algorithm by updating the model and the parameter under the model in each iteration. Specifically, the EMS algorithm performs expectation step (E-step) and model selection step (MS-step) alternately to find the minimum point of the observed generalized information criteria (GIC). In E-step, it computes the $Q$ function (i.e., the expectation of complete information criteria) for each candidate model while in MS-step it selects an optimal model with the minimum value of the $Q$ function.

There are two drawbacks of EMS. First, \citet{JiangNR2015} proved the global convergence of EMS depending on two strong assumptions. One assumption is that the points at which EMS terminates is a subset of the minimum points of GIC. Another assumption is that GIC has a unique minimum point. Second, in practice it may not be computationally feasible to perform the MS-step, especially for high dimensional data.
For example, there will be a total of $2^p$ possible models for a linear regression problem with $p$ covariates and it is even difficult to  find an optimal model with the minimum value of the $Q$ function for moderate $p$.

In this paper, we develop a more simple but feasible method which on the contrary seeks only a decrease in the $Q$ function value in the MS-step, which in turn leads to a decrease in the observed GIC. The resulting method is called the generalized EMS (GEMS) algorithm, which includes the EMS algorithm as a special case. We obtain several numerical convergence results of the GEMS algorithm. As a special convergence result, we present that EMS will converge without above two assumptions. This property of EMS will borad its applications.

The rest of this paper is organized as follows. Section 2 gives some necessary notations and review of the existing EMS algorithm. In Section 3, we present some numerical convergence results and useful convergence properties of the EMS algorithm. In section 4, we apply GEMS for variable selection for generalized linear model with mixed predictors. In Section 5, we apply the GEMS algorithm for Gaussian graphical model selection. In Section 6, we conduct numerical experiments to compare GEMS with existing competitors. We illustrate our GEMS algorithm for three real data sets in Section 7. Finally, we draw our conclusion in Section 8. In Appendix, we prove our main results.

\section{A Review of the EMS algorithm}
In this section, we first give some necessary notations about incomplete data and then review the existing  EMS algorithm for model selection.


Let $\mathcal{M}$ be the model space (i.e., the set of all possible candidate models) and let $\Theta_{M}$ the parameter space under the model $M\in\mathcal{M}$. If $m(M, \theta_M, Y)$ represents a measure of lack-of-fit based on the complete data $Y$, the generalized information criteria (GIC) is defined as
\begin{eqnarray}
c(M, \theta_M, Y) = m(M, \theta_M, Y) + p(M),\label{gic}
\end{eqnarray}
where $\theta_M\in \Theta_M$ and $p(\cdot)$ is a penalty function on the complexity of $M$. When there are missing data, one can define the observed GIC as
\begin{eqnarray}
g(M, \theta_M, Y_o) = m(M, \theta_M, Y_o) + p(M),\label{obsgic}
\end{eqnarray}
where $m(M, \theta_M, Y_o)$ is the observed measure of lack-of-fit based on the observed data, $Y_o$. \citet{IbrahimZT2008} and \citet{JiangNR2015} proposed to choose the model that minimizes the observed $g(M, \theta_M, Y_o)$ in (\ref{obsgic}) for model selection. However, there is generally no closed-form solution if $m(M, \theta_M, Y_o)$ is taken to be minus twice the observed log-likelihood. To overcome this issue, \citet{IbrahimZT2008} proposed the IC$_{QH}$ information by computing the approximation of $g(M, \theta_M, Y_o)$ via the EM algorithm. \citet{JiangNR2015} generalized the well-known EM algorithm to the so-called EMS algorithm which extends the concept of parameters to include both the model and the parameters under the model. That is, it defines the new parameter to be $\psi = (M, \theta_M)$ for $M\in\mathcal{M}$ and $\theta_M\in\Theta_M$ and let $\Psi$  be the new parameter space.

Before describing the EMS algorithm, we define the new $Q$ and $H$ functions which are inspired by the EM algorithm as follows. For any parameter $\psi$ and $\tilde\psi$, define $Q(\psi; \tilde\psi) = E(c(\psi, Y) \mid  Y_o, \tilde\psi)$ and $H(\psi; \tilde\psi) = E(m(\psi, Y) - m(\psi, Y_o) \mid Y_o, \tilde\psi)$. Hence, we have
\begin{eqnarray}
g(\psi, Y_o) & = & E(g(\psi, Y_o) \mid  Y_o, \tilde\psi) 
\nonumber\\
& = & E(m(\psi, Y) + p(M) \mid  Y_o, \tilde\psi) - E(m(\psi, Y) - m(\psi, Y_o) \mid Y_o, \tilde\psi) \nonumber\\
& = & Q(\psi; \tilde\psi) - H(\psi; \tilde\psi).\label{gexp}
\end{eqnarray}

The EMS first chooses an initial parameter $\psi^{(0)} = (M^{(0)}, \theta^{(0)})$ such that $g(M^{(0)}, \theta^{(0)}, Y_o) < +\infty$, where $M^{(0)}$ is an initial model and $\theta^{(0)}\in\Theta_{M^{(0)}}$ is an initial parameter. It aims to find an optimal parameter $\psi^* = (M^*, \theta^*)$ which yields the minimum of the observed GIC in (\ref{obsgic}) by iteratively applying the Expectation step (E-step) and the Model Selection step (MS-step) until some convergence criteria is satisfied. Specifically, we define the EMS iteration $\psi^{(t)} \rightarrow \psi^{(t+1)}\in \mathbf A(\psi^{(t)})$ as follows:
\begin{adjustwidth}{1cm}{0cm}
\begin{itemize}
\item[E-step.] Compute $Q(\psi; \psi^{(t)}) = E(c(\psi, Y) \mid Y_o, \psi^{(t)})$ for each parameter $\psi = (M, \theta_M)\in\Psi$.

\item[MS-step.] Choose $\psi^{(t+1)}$ to be any value in $\mathbf A(\psi^{(t)}) = \arg\min_{\psi'\in\Psi}Q(\psi';\psi^{(t)})$, i.e., the set of minimum points of $Q(\psi';\psi^{(t)})$ over $\Psi$.
\end{itemize}
\end{adjustwidth}
Note that $\mathbf A(\psi) = \arg\min_{\psi'} Q(\psi'; \psi)$ for any $\psi\in\Psi$. In fact, $\mathbf A$ is a point-to-set mapping that assigns to every point $\psi\in \Psi$ a subset of $\Psi$. 
To choose $\psi^{(t+1)}$ in the MS-step, we first choose $\hat\theta^{(t+1)}_M \in \Theta_M$ which minimizes $Q(M, \theta_M; \psi^{(t)})$ for each model $M\in\mathcal{M}$.
We then select the optimal model $M^{(t+1)}$ which minimizes $Q(M, \hat\theta^{(t+1)}_M; \psi^{(t)})$. Thus, we get $\psi^{(t+1)} = (M^{(t+1)}, \hat\theta^{(t+1)}_{M^{(t+1)}})$ which leads to $g(\psi^{(t+1)}, Y_o) \leq g(\psi^{(t)}, Y_o)$.

Before we present the convergence of the EMS algorithm, we first define $\Psi_0 = \arg\min_{\psi} \{g(\psi, Y_o)\}$ and $\Psi_1 = \{\psi |Q(\psi; \psi)\leq Q(\psi'; \psi)$ for any $\psi'\in \Psi\}$. Note that, $\Psi_0$ is the set of minimum points of the observed GIC $g(\psi, Y_o)$) and $\Psi_1$ is the set of stationary points at which the EMS stops decreasing. Moreover, if we fix the model in $\Psi_1$, that is, $M = M'$ for $\psi = (M, \theta_M)$ and $\psi' = (M', {\theta'}_{M'})$, then $\Psi_1 = \{(M, \theta_M) |Q((M, \theta_M); (M, \theta_M))\leq Q((M', {\theta'}_{M'}); (M, \theta))$ for any ${\theta'}_{M'} \in \Theta_{M'}\}$ will reduce to a parameter space $\Theta_1 = \{\theta_M| E(m(M, \theta_M, Y) | Y_o, (M, \theta_M)) \leq E(m(M, \theta'_M, Y) | Y_o, (M, \theta_M))$ for any ${\theta'}_{M}\in \Theta_{M}\}$. In fact, when the measure $m$ of lack-of-fit is taken to be minus twice log-likelihood, $\Theta_1$ is the set of stationary points in the EM algorithm, see details Section 3.3 of \citet{McLachlanKrishnan2007}.

\citet{JiangNR2015} obtained the global convergence of the EMS algorithm under the following assumptions.
\begin{enumerate}
\item[A1.] The model space $\mathcal M$ is finite and the parameter space $\Theta_M$ is compact for any $M\in\mathcal M$.

\item[A2.]  For any fixed $M_j \in\mathcal M, j = 0, 1$, as $\theta_j, \tilde\theta_j \in\Theta_{M_j}$ and $\tilde\theta_j \rightarrow \theta_j, j =0, 1$, we have $E\{m(M_1,  \tilde\theta_1, Y) - m(M_1, \theta_1, Y)|y_o, M_0, \tilde\theta_0\}\rightarrow 0$ and\\ $E\{m(\psi_1, Y) | y_o, M_0, \tilde\theta_0\} - E\{m(\psi_1, Y) | y_o, M_0, \theta_0\}\rightarrow 0$.

\item[A3.] For any parameters $\psi$ and $\tilde\psi$, we have  $H(\tilde\psi; \tilde\psi) \leq H(\psi; \tilde\psi)$, i.e.,
\begin{eqnarray}
E(m(\tilde\psi, Y) - m(\tilde\psi, Y_o) \mid Y_o, \tilde\psi) \leq E(m(\psi, Y) - m(\psi, Y_o) \mid Y_o, \tilde\psi).
\end{eqnarray}

\item[A4.] $\{\Psi\setminus \Psi_0\} \cap \Psi_1 = \emptyset$;

\item[A5.] $|\Psi_0| = 1$ where $|\cdot|$ denotes cardinality.
\end{enumerate}

There are two drawbacks of the EMS. First, the global convergence of EMS depends on strong assumptions A4 and A5. In fact, Assumption A4 implies $\Psi_1\subseteq \Psi_0$, that is, the EMS must stops at some minimum point. 
Assumption A5 means the minimum point in $\Psi_0$ is unique. However, A4 and A5 are hard to be verified in practice. 
Second, it sometimes may not be computationally feasible to perform the MS step, especially for high dimensional data. In addition, it is not necessary that $\psi^{(t+1)}$ actually minimizes the $Q$ function for the observed GIC $g(\psi, Y_o)$.

\section{The GEMS algorithm and its convergence properties}
Inspired by the idea of the generalized EM defined by \citet{DempsterLR1977}, in which one chooses a value that increases the $Q$ function for M-step of EM (see \citet{McLachlanKrishnan2007}), we propose a generalized EMS (GEMS) algorithm in this section. For MS-step of GEMS, we choose a value $\psi^{(t+1)}$ such that $Q(\psi; \psi^{(t)})$ will decrease (i.e., $Q(\psi^{(t+1)}; \psi^{(t)}) \leq Q(\psi^{(t)}; \psi^{(t)})$), rather than minimize $Q(\psi; \psi^{(t)})$. It should be noticed that the point-to-set mapping in GEMS is $\mathbf A(\psi) = \{\tilde\psi | Q(\tilde\psi; \psi) \leq Q(\psi; \psi)\}$. Obviously, EMS is a special case of GEMS.

In this paper, we assume the following regularity conditions:
\begin{enumerate}
\item[C1.] The model space $\mathcal M$ is finite and the parameter space $\Theta_M$ is compact for any $M\in\mathcal M$.

\item[C2.] (\romannumeral 1) For any fixed $M\in\mathcal M$, $g(M, \theta_M, Y_o)$ is continuous in $\Theta_M$.

        (\romannumeral 2) For $\psi^{(t)}_0\rightarrow \psi_0$ and $\psi^{(t)}_1\rightarrow \psi_1$, we have $Q(\psi^{(t)}_1; \psi^{(t)}_0) - Q(\psi_1; \psi^{(t)}_0) \rightarrow 0$ and $Q(\psi_1; \psi^{(t)}_0) \rightarrow Q(\psi_1; \psi_0)$.
\item[C3.] For any parameters $\psi$ and $\tilde\psi$, we have  $H(\tilde\psi; \tilde\psi) \leq H(\psi; \tilde\psi)$, i.e.,
\begin{eqnarray}
E(m(\tilde\psi, Y) - m(\tilde\psi, Y_o) \mid Y_o, \tilde\psi) \leq E(m(\psi, Y) - m(\psi, Y_o) \mid Y_o, \tilde\psi).
\end{eqnarray}
\end{enumerate}
Here, $\psi^{(t)}_0 = (M^{(t)}_0, \theta^{(t)}_0)\rightarrow \psi_0 = (M_0, \theta_0)$ means that for any $\epsilon>0$ there exists $T>0$ such that $M^{(t)}_0 = M$ and $|\theta^{(t)}_0 - \theta_0| <\epsilon$ for any $t>T$.

Note that conditions C1 and C3 are the same as Assumptions A1 and A3 in \citet{JiangNR2015}, respectively. Condition C2 (\romannumeral 1) was used in the proof in \citet{JiangNR2015} although it was not listed as one of the assumptions. Under condition C1, it is not difficult to show that condition C2 (\romannumeral 2) is equivalent to Assumption A2 in \citet{JiangNR2015}. In fact, condition C2 (\romannumeral 2) follows from the fact  that $Q(M_1, \theta_1; M_0, \theta_0)$ is continuous in both $\theta_1$ and $\theta_0$ for fixed $M_0$ and $M_1$. If the negative log-likelihood function is taken as a measure of lack-of-fit, this continuity of $Q(M_1, \theta_1; M_0, \theta_0)$ is the same continuity (i.e., (10)) of $Q$ function in the EM being required in \citet{Wu1983} to prove that the limit points of the sequence generated by EM are stationary points of the observed log-likelihood. As pointed out by \citet{Wu1983}, the curved exponential family and many other densities outside the exponential family satisfy this continuity requirement. \citet{McLachlanKrishnan2007} showed that this continuity is very weak and should hold in most practical situations. Therefore, condition C2 (\romannumeral 2) holds in most applications.

Most importantly, it should be noticed that conditions C1 to C3 are commonly required by convergence of the EM algorithm if the model $M$ is fixed and the measure $m$ of lack-of-fit is taken to be minus twice log-likelihood. See more in Chapter 3 of \citet{McLachlanKrishnan2007}. In the paper, we require these mild conditions to prove the convergence of GEMS and EMS.


First, we present a necessary condition (i.e., inequality (\ref{nece})) that all minimum points of $g(\psi, Y_o)$ should satisfy.

\begin{theorem}\label{neceCond} Under condition C3, we have $\Psi_0\subseteq \Psi_1$; i.e., for any $\psi^*\in \Psi_0$ we have
\begin{eqnarray}
Q(\psi^*; \psi^*) \leq Q(\psi; \psi^*) \mbox{ for any $\psi\in\Psi$.}\label{nece}
\end{eqnarray}
\end{theorem}

In fact, inequality (\ref{nece}) generalizes (3.11) in Section 3.2 of \citet{McLachlanKrishnan2007} for the EM algorithm. From  inequality (\ref{nece}), the EMS algorithm satisfies the self-consistency property; i.e, for a minimum point $\psi^*\in\Psi_0$, the EMS algorithm can choose $\psi^*$ as a minimum point of $Q(\psi; \psi^*)$.
Next, we show some nice properties of GEMS.

\begin{theorem}\label{non-increasing}
If condition C3 holds, for every GEMS with mapping $\mathbf A$  we have
\begin{eqnarray}
g(\psi, Y_o) \geq g(\tilde\psi, Y_o) \mbox{ for any } \psi\in\Psi \mbox{ and } \tilde\psi\in \mathbf A(\psi),
\end{eqnarray}
where the equality holds if and only if  $Q(\psi; \psi) = Q(\tilde\psi; \psi)$ and $H(\psi; \psi) = H(\tilde\psi; \psi)$.
\end{theorem}

It follows from Theorem \ref{non-increasing} that every GEMS generates a non-increasing sequence $\{g(\psi^{(t)}, Y_o), t = 0, 1, \dots\}$. Furthermore, under conditions C1 to C3,  any sequence $\{g(\psi^{(t)}, Y_o), t = 0, 1, \dots\}$ generated by GEMS is bounded, therefore $\{g(\psi^{(t)}, Y_o), t = 0, 1, \dots\}$ is convergent to some value $g^*$. We also get the following corollary.

\begin{corollary}
For $\psi^*\in\Psi_0$ and any GEMS with mapping $\mathbf A$, we have (\romannumeral 1) for any $\psi'\in\mathbf A(\psi^*)$, $g(\psi', Y_o) = g(\psi^*, Y_o)$ and $Q(\psi'; \psi^*) = Q(\psi^*; \psi^*)$; (\romannumeral 2) $|\Psi_0| = 1$ implies that $\mathbf A(\psi^*) = \{\psi^*\}$.
\end{corollary}

Next, we describe the convergence properties of GEMS and EMS.

\begin{theorem}\label{gemsC} Assume that
\begin{eqnarray} {\mbox if }\psi\not\in \Psi_1 {\ \mbox (and}\ \Psi_0, {\mbox respectively)\ then\ } g(\tilde\psi, Y_o) < g(\psi, Y_o) {\mbox\ for\ all\ } \tilde\psi\in \mathbf A(\psi)\label{des}
\end{eqnarray}
 for any GEMS algorithm with mapping $\mathbf A$. Under conditions C1, C2 and C3, we have all limit points of $\{\psi^{(t)}; t = 0, 1, \dots\}$ generated by the GEMS algorithm belong to $\Psi_1$ (and $\Psi_0$, respectively).
\end{theorem}

\begin{corollary}\label{emsC1} Let
$\{\psi^{(t)}; t = 0, 1, \dots\}$ be the sequence generated by the EMS algorithm with mapping $\mathbf A$. Under conditions C1, C2 and C3, all limit points of $\{\psi^{(t)}\}$ belong to $\Psi_1$. 
\end{corollary}

It follows from Theorem \ref{neceCond} and Corollary \ref{emsC1} that any limit point of the sequence generated by the EMS algorithm satisfies  inequality (\ref{nece}); i.e., the necessary condition of being minimum point of $g(\psi, Y_o)$. Corollary \ref{emsC1} does not rely on two strong assumptions A4 and A5 of \citet{JiangNR2015}, therefore it is an attractive complementary of the global convergence proposed by \citet{JiangNR2015}.
Theorem \ref{gemsC} and Corollary \ref{emsC1} are more useful from the user point of view, because conditions C1, C2 and C3 readily hold for many models.

\section{Variable selection for generalized linear model with mixed missing data}
Suppose that $y$ is the response variable and $w = (w_1, \dots, w_q)$ is a q-dimensional categorical predictor and $z = (z_1, \dots z_p)$ is a p-dimensional metric predictor. We first assume categorical predictors $w_j$ have values $w_j\in \{0, 1, \dots, k_j\}$ for $j = 1, 2, \dots, q$.
To include the categorical predictors into generalized linear models, we could use dummy variables defined by $x_{jr} = 1$ if $w_j = r$ and $x_{jr} = 0$ otherwise. Therefore, we yield the linear predictor
\begin{eqnarray}
\eta &=& \beta_0 + \sum_{j=1}^q\sum_{r=0}^{k_j}x_{jr}\beta_{jr} + \sum_{\gamma=1}^p x_{\gamma+q}\beta_{\gamma+q} \nonumber \\
     &=& \beta_0 + \sum_{j=1}^q x_j^\mathsf{T}\beta_j + \sum_{\gamma=1}^p x_{\gamma+q}\beta_{\gamma+q}\label{predictor}
\end{eqnarray}
where $x_j = (x_{j0}, x_{j1}, \dots, x_{jk_j})^\mathsf{T}$, $\beta_j = (\beta_{j0}, \beta_{j1}, \dots, \beta_{jk_j})^\mathsf{T}$ collects all parameters linked to predictor $w_j$ for $j=1, \dots, p$ and $\beta_{\gamma+q}$ is a parameter of $x_{\gamma+q} = z_\gamma$ for $\gamma = 1, \dots, p$. For means of identifiability, we set $\beta_{j0} = 0$ for $j=1, \dots, q$. Thus $x_j$ and $\beta_j$ can be reduced to $x_j = (x_{j1}, \dots, x_{jk_j})^\mathsf{T}$, $\beta_j = (\beta_{j1}, \dots, \beta_{jk_j})^\mathsf{T}$. Then we get a predictor $x = (x_1^\mathsf{T}, \dots, x_q^\mathsf{T}, x_{q+1}, \dots, x_{q+p})^\mathsf{T}$ involving dummy variables and parameter $\beta = (\beta_1^\mathsf{T}, \dots, \beta_q^\mathsf{T}, \beta_{q+1}, \dots, \beta_{q+p})^\mathsf{T}$, thus (\ref{predictor}) can be simplified to
$\eta = \beta_0 + x^\mathsf{T}\beta\label{predictor2}$.

The joint density function is given by
\begin{eqnarray}
f(y, x|\tau, \alpha, \beta) = f(y|x, \tau, \beta)f(x|\alpha),
\end{eqnarray}
where $f(y|x, \tau, \beta)$ denotes the density of $y$ given $x$ and $f(x|\alpha)$ denotes the density of $x$. For generalized linear models, we assume that $f(y|x, \tau, \beta)$ satisfies
\begin{eqnarray}
E(y| x; \tau, \alpha, \beta) = g(\beta_0 + x^\mathsf{T}\beta)\label{glm}
\end{eqnarray}
where $g(\cdot)$ is a known link function and $\tau$ denotes the additional parameters (see, e.g.,  \citet{McCullaghNelder1989}).

In this section we consider variable selection problem in generalized linear models. If there is no missing value, smart stepwise procedures (such as step in R) can add or drop dummy variables corresponding to the same categorical variable at a time for low dimensional settings. For high dimensional settings, the group lasso is a natural and computationally convenient approach to select predictors \citep{YuanLin2006}. 

When missing data is present, there are three major types of variable selection methods. They are namely the likelihood-based method (see, e.g., \citet{HortonLaird1999}, \citet{GarciaIZ2010a, GarciaIZ2010b}, \citet{StadlerBuhlmann2012}, \citet{SabbeTO2013}), inverse probability weighting method (see, \citet{JohnsonLZ2008}) and multiple imputation method (see, e.g., \citet{LongJohnson2015}, \citet{LiuWFW2016} and \citet{ZhaoL2017}). In above mentioned papers, only \citet{SabbeTO2013} considered variable selection when there are categorical and continuous predictors. In this second, we focus on those likelihood-based methods by the GEMS algorithm.

\subsection{Strong decomposition tree model for modeling predictors}\label{pgm}
An important issue with missing predictor data is the specification of a parametric model for missing predictors. \citet{IbrahimLC1999} modeled the joint distribution of the predictors as a product of one-dimensional parametric conditional distributions. They wrote the joint distribution of the (p+q)-dimensional predictor vector as 
\begin{eqnarray}
p(w, z)&=& p(z_1 |\alpha_1)p(z_2 |z_1, \alpha_2)\dots p(z_p |z_1, \dots, z_{p-1}, \alpha_p)\nonumber\\
 && p(w_1 |z, \alpha_{p+1})p(w_2 |z, w_1, \alpha_{p+2})\dots p(w_q | z, w_1, \dots, w_{q-1}, \alpha_{p+q})
\end{eqnarray}
where $\alpha_k$ is a vector of indexing parameters for the $k$th conditional distribution and $\alpha = (\alpha_1, \dots, \alpha_{p+q})$.
They suggested to specify one-dimensional (or joint) distributions for the continuous predictors first and then to obtain the one-dimensional distributions for the categorical predictors by conditioning on the continuous predictors. For example, they assume continuous predictor $z_k$ follows from a normal distribution given $z_1, \cdots, z_{k-1}$ and they specify a logistic regression model for the categorical predictor $w_j$ with $z_1, \cdots, z_p$ and $w_1, \cdots, w_{j-1}$ as predictors.
However,  the missing pattern maybe changes from one observation to anther in many applications. Therefore, it is difficult to specify the variables enter order.

\citet{SabbeTO2013} modeled the predictors by the general location model (GLoMo). In the unrestricted GLoMo, the cells of the marginal contingency table of the categorical variables are modeled as a single multinomial vector-valued variable with cell probabilities $\pi_c$. In short, each cell represents one unique combination of all categorical variables. Conditional on the categorical variables, that is, on the cell $c$, the continuous variables are assumed to have a multivariate normal distribution $N(\mu_c, \Sigma)$ with the mean depending on cell $c$. That is, the predictors follow jointly a conditional Gaussian distribution. More details can be found in \citet{Schafer1997}. However, the maximum likelihood estimation of unrestricted GLoMo exists if and only if the number of observations in each cell is greater than the number of continuous variables according to Proposition 6.9 in \citet{Lauritzen1996}. 
Thus, the unrestricted GLoMo may not be estimated for high dimensional setting when the sample size is smaller than the number of predictors. \citet{Schafer1997} suggested restricted GLoMo in which they modeled the categorical predictors by log-linear model and specify the relationship of the continuous predictors to the categorical
ones. But restricted GLoMo depends on the prior knowledge of the correlation structure among the predictors which is rarely available in practice.

\begin{figure}[t]\centering
\epsfig{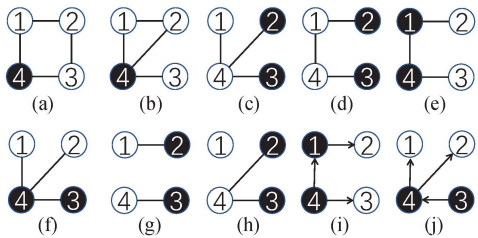}
\caption{Continuous variables are represented as circles, and categorical variables as dots. (a) is not a tree because it contains 1, 2, 3, 4, 1 as a cycle. (b) contains a cycle 1, 2, 4, 1, so it is not a tree. (c) to (f) are trees. (c) contains a forbidden path 2, 4, 3 and (d) contains a forbidden path 2, 1, 4, 3, so (c) and (d) are not SD-trees. (e) and (f) do not contain any forbidden path, so they are SD-trees. (g) and (h) are forests, and (g) is a SD-forest but (h) is not a SD-forest. (i) is a single parent DAG with 4 as its root and (j) is a single parent DAG with 3 as its root}\label{TreeForest}
\end{figure}

In this paper, we model the mixed predictors by a strong decomposition tree model or a strong decomposition forest model. Here we give simple brief definitions. For more details, please referee to \citet{EdwardsAbreuLabouriau2010} and \citet{AbreuEdwardsLabouriau2010}. A tree  is a connected graph $(V, E)$ without cycles, where $V$ is the vertex set and $E$ is the edge set. A vertex $v\in V$ associate with the random variable $X_v$. A tree is called a strong decomposition tree (SD-tree) if it contains no forbidden paths. A forbidden path is a path between two non-adjacent categorical vertices passing through only continuous vertices. 
A strong decomposition forest (SD-forest) is a forest in which each connected component is a SD-tree. See examples in Figure \ref{TreeForest}.

We assume that the mixed predictors $x = (w, z)$ follow a conditional Gaussian (CG) distribution which satisfies global Markov property corresponding to a SD-tree or a SD-forest. That is, if $u$ and $v$ are not adjacent in the SD-tree or the SD-forest, then $x_u$ and $x_v$ are conditional independent given all other variables. For example, $x_2\ind x_4 | x_1, x_3$, $x_2\ind x_3|x_1, x_4$, in Figure \ref{TreeForest} (e). Therefore, SD-tree or SD-forest could encodes sets of conditional independence relations among mixed predictors.

There are at least three advantages of SD-tree (SD-forest) model. First, maximum likelihood estimation exists usually for real data sets.  According to Markov property, we have the probability densities of such models can be factorized as
\begin{eqnarray}
f(x) = \prod_{v\in V}f(x_v)\prod_{(u, v)\in E} \frac{f(x_u, x_v)}{f(x_u)f(x_v)}, \label{factorization}
\end{eqnarray}
where $(u, v)\in E$ is an edge and $f(x_u, x_v)$ and $f(x_u)$ are the marginal probability densities. That is, these models can be decomposed into a series of marginal models on two variables $x_u$ and $x_v$ of an edge $(u, v)\in E$ and the marginal models on each variable $x_u$. The maximum likelihood estimation exists if and only if there exist maximum likelihood estimations for all marginal models on all edges $(u, v)\in E$ and on all variables $u\in V$. There are five types of marginal models in (\ref{factorization}). (\romannumeral1) For models on two categorical variables $x_u, x_v$, there exists maximum likelihood estimation if and only if the number of observations in each cell of $x_u, x_v$ is positive. (\romannumeral2) For models on two continuous variables, there exists maximum likelihood estimation if and only if the sample size is greater than 2. (\romannumeral3) For models with one categorical variable and one continuous variable, there exists if and only if there are at least two observations in each cell of the categorical variable. (\romannumeral4) For marginal model on one categorical variables, maximum likelihood estimation exists if and only if the number of observations in each cell of this categorical variable is positive. (\romannumeral5) For marginal model on one continuous variable, maximum likelihood estimation exists if and only if the sample size is greater than 1. Therefore, the condition that maximum likelihood estimate of the SD-tree (SD-forest) exist is usually satisfied for real data sets. 
Furthermore, these models have an explicit formula for the maximum likelihood estimation as shown in Chapter 6 of \citet{Lauritzen1996}.

Second, we could learn a SD-tree (SD-forest) from the data without any prior knowledge or expert knowledge.
To find a SD-tree (SD-forest) among variables with maximum likelihood estimates or with minimal BIC, \citet{EdwardsAbreuLabouriau2010} proposed a method as follows. First, they define the sample mutual information $I_{u,v}$ or the BIC penalized mutual information $I_{u,v}^{BIC} = I_{u,v} - \ln(n)k_{u,v}/2$ for each pair of variables $x_u, x_v$, where $k_{u,v}$ is the number of free parameters associated with $I_{u,v}$. Note that, for a SD-tree $T = (V, E)$, the maximized log-likelihood is $\sum_{(u,v)\in E} I_{u,v}$ and the BIC information is $-2\sum_{(u,v)\in E}I_{u,v}^{BIC}$ as shown by \citet{EdwardsAbreuLabouriau2010}. If $I_{u,v}$ or $I_{u,v}^{BIC}$ is viewed as edge weights on the complete graph (a simple undirected graph in which every pair of distinct vertices is connected by a unique edge) with vertex set $x = (w, z)$, then they efficiently obtain the maximum likelihood tree or minimal BIC SD-tree (SD-forest) by maximum spanning tree algorithm (for example, Kruskal's algorithm with  $O((p+q)^2 \ln(p+q))$ time complexity). Therefore, the SD-tree (SD-forest) is attractive for high-dimensional setting, because we could find efficiently the SD-tree (SD-forest) with minimal BIC.

Third, we could draw samples from the posterior distribution of missing predictors given some observed values based on a SD-tree model. In fact, a SD-tree is Markov equivalent to a single-parent directed acyclic graph (DAG).
As inspired by \citet{EdwardsAbreuLabouriau2010}, we first find one categorical variable $x_R$ as a vertex with no parent. Then we orient all edges in the SD-tree away from the categorical variable $x_R$, thus we will get a DAG $\vec{G}$ with the restriction that continuous vertices are not allowed to point to any categorical vertex. If a vertex $u$ points to a vertex $v$ in the DAG $\vec{G}$, then $u$ is a parent vertex of $v$. 
Since each vertex has at most one parent vertex in obtained $\vec{G}$, such DAG is called a single-parent DAG. For example, the SD-trees in Figure 1 (e) is Markov equivalent to the DAG with $4$ as a root in Figure 1 (i), and the SD-tree in Figure 1 (f) is Markov equivalent to the DAG with $3$ as a root in Figure 1 (j). For each directed edge with a parent vertex $u$ pointing to a child vertex $v$, we can derive a conditional density $f(x_v|x_u)$ of $x_v$ given $x_u$ from the joint density $f(x)$. Thus, $f(x)$ can be factorized into another form as
\begin{eqnarray}
f(x) = \prod_{v\in V}f(x_v|x_u), \label{factorization}
\end{eqnarray}
where $u$ is a parent vertex of $v$ in $\vec{G}$. Thus we get a pair $(\vec{G}, f(x))$ called mixed Bayesian network. For mixed Bayesian network, local propagation algorithms of \citet{Cowell2005} and \citet{LauritzenJensen2001} could be used to compute posterior distributions given some observed values. Furthermore, local method such as Algorithm 7.1 of \citet{Cowell2005} could draw sample from posterior distributions. In this paper, we write C codes to implement Algorithms 7.1 of \citet{Cowell2005} for sampling.

\subsection{Variable selection with mixed missing data by GEMS}
Let $y_i$, $x_i$ be the $i$ th independent and identically distributed realisation of $y$ and $x$. We allow missing data in predictors. For $i$ th observation, let $x_{i,obs_i}$ and $x_{i,mis_i}$ be the observed and missing components. Thus, the observed data is $Y=(y_i)_{i=1, \dots, n}$ and $X_{obs} = (x_{i,obs_i})_{i=1, \dots, n}$.

Here, we focus on the sparse estimates for the regression coefficients $\beta$ in generalized linear model (\ref{glm}).
In GEMS, we adopt the Bayesian information criteria $c(M, \theta_M, Y, X) = 2 l_{Y, X} + \ln(n)(\mathrm{df}_\beta + \mathrm{df}_\alpha)$, where $\mathrm{df}_\beta$ and $\mathrm{df}_\alpha$ are the numbers of non-zero parameters in $\beta$ and $\alpha$ respectively, and the negative log-joint likelihood $l_{Y,X}=l_{Y|X} + l_X$. Here, $l_X = -\sum_{i=1}^n \log f(x_i|\alpha)$ is the negative log-marginal likelihood of $X$ and $l_{Y|X} =  -\sum_{i=1}^n\log f(y_i|x_i, \tau, \beta)$ is the negative log-conditional likelihood of $Y$ given $X$.

Suppose we have $\psi^{(t)} = (M^{(t)}, \theta^{(t)})$ in the $t$ th iteration, where $\theta^{(t)} = (\alpha^{(t)}, \tau^{(t)}, \beta^{(t)})$. 
We have the $Q$ function can be divided into two parts
\begin{eqnarray}
Q(\psi; \psi^{(t)}) = Q_1(\alpha; \psi^{(t)}) + Q_2(\tau, \beta; \psi^{(t)})
\end{eqnarray}
where $Q_1(\alpha; \psi^{(t)}) = E(2 l_X \mid Y, X_{obs}, \psi^{(t)}) + \ln(n) \mathrm{df}_\alpha$ and $Q_2(\tau, \beta; \psi^{(t)}) = E(2 l_{Y|X}  \mid Y, X_{obs}, \psi^{(t)} ) + \ln(n) \mathrm{df}_\beta$. Let
\begin{eqnarray}
Q_{1i}(\alpha; \psi^{(t)}) &=& -2E(\log f(x_{i,obs_i}, x_{i,mis_i})   \mid  y_i, x_{i,obs_i}, \psi^{(t)}) + n^{-1}\ln(n) \mathrm{df}_\alpha,\nonumber\\
Q_{2i}(\tau, \beta; \psi^{(t)}) &=& -2E(\log f(y_i|x_{i,obs_i}, x_{i,mis_i})  \mid y_i, x_{i,obs_i}, \psi^{(t)}) + n^{-1}\ln(n) \mathrm{df}_\beta,\nonumber
\end{eqnarray}
then $Q_1(\alpha; \psi^{(t)}) = \sum_{i=1}^n Q_{1i}(\alpha; \psi^{(t)})$ and $Q_2(\tau, \beta; \psi^{(t)}) = \sum_{i=1}^n Q_{2i}(\tau, \beta; \psi^{(t)})$. Hence, we need to calculate each $Q_{1i}$ and $Q_{2i}$.

However, the expectation may be difficult to calculate $Q_{1i}$ and $Q_{2i}$. \citet{WeiTanner1990} approximated the expectation in a classical Monte Carlo way. If the density of the response variable $Y$ as a function of the mean of $Y$ is bounded above by a known constant $c$, then
\begin{eqnarray}
f(x_{i,mis_i} | y_i, x_{i,obs_{i}}, \psi^{(t)})
&\propto& f(y_i | x_i, \tau^{(t)}, \beta^{(t)})f(x_{i,mis_i} | x_{i,obs_i}, \alpha^{(t)})\nonumber\\
&\leq& cf(x_{i,mis_i} | x_{i,obs_i}, \alpha^{(t)}).\nonumber
\end{eqnarray}
Since Algorithm 7.1 of \citet{Cowell2005} can be applied to generate samples from $f(x_{i,mis_i} | x_{i,obs_i}, \alpha^{(t)})$ according to SD-tree (SD-forest), we could use the acceptance-rejection method to draw samples from $f(x_{i,mis_i} | y_i, x_{i,obs_{i}}, \psi^{(t)})$. If the bound $c$ of the density of $Y$ is not available, we could generate samples from $f(x_{i,mis_i}  | y_i, x_{i,obs_{i}}, \psi^{(t)})$  by Gibbs sampler, in which the adaptive rejection algorithm of \citet{GilksWild1992} for continuous is applied due to log-concave property of $f(y_i | x_i, \tau^{(t)}, \beta^{(t)})$ and $f(x_i | \alpha^{(t)})$ in the components of $x_i$.
For more details, one can refer to \citet{IbrahimLC1999}. For the $i$-th observation, $m$ independent realisations of $x_{i,mis_i}$ are drawn from $f(x_{i,mis_i}  \mid y_i, x_{i,obs_{i}}, \psi^{(t)})$, say $x_{i,mis_i}^{(1)}, \dots, x_{i,mis_i}^{(m)}$. Therefore,
\begin{eqnarray}
Q_{1i}(\alpha; \psi^{(t)}) \approx - \frac{1}{m}\sum_{h=1}^{m} 2\log f(x_{i,mis_i}^{(h)}, x_{i,obs_i}  | \alpha) + \frac{\ln(n)}{n} \mathrm{df}_\alpha, \nonumber
\end{eqnarray}
and
\begin{eqnarray}
Q_{2i}(\tau, \beta; \psi^{(t)}) \approx - \frac{1}{m}\sum_{h=1}^{m} 2\log f(y_i \mid x_{i,mis_i}^{(h)}, x_{i,obs_i},  \tau, \beta) + \frac{\ln(n)}{n} \mathrm{df}_\beta. \nonumber
\end{eqnarray}
Hence, the $Q$ function is approximated to
$$Q(\psi; \psi^{(t)}) \approx \frac{\tilde{Q}_1(\alpha; \psi^{(t)}) + \tilde{Q}_2(\tau, \beta; \psi^{(t)})}{m}$$
where
\begin{eqnarray}
\tilde{Q}_1(\alpha; \psi^{(t)}) &=& -\sum_{i=1}^n \sum_{h=1}^{m} 2\log f(x_{i,mis_i}^{(h)}, x_{i,obs_i}  | \alpha) + m\ln(n) \mathrm{df}_\alpha  \label{aq1}\\
\tilde{Q}_2(\tau, \beta; \psi^{(t)}) &=& - \sum_{i=1}^n \sum_{h=1}^{m} 2\log f(y_i \mid x_{i,mis_i}^{(h)}, x_{i,obs_i}, \tau, \beta) + m\ln(n)\mathrm{df}_\beta\label{aq2}
\end{eqnarray}

For MS-step of GEMS, we choose a value $\psi^{(t+1)}$ such that $Q(\psi; \psi^{(t)})$ decreases (i.e., $Q(\psi^{(t+1)}; \psi^{(t)}) \leq Q(\psi^{(t)}; \psi^{(t)})$) by the following heuristic method. First, we minimize $\tilde{Q}_1$ over SD-forests. In particular, we define the penalized mutual information quantity $I_{u,v}^{mBIC} = I_{u,v} - m\ln(n)k_{u,v}/2$ for each pair of variables $x_u, x_v$ according to (\ref{aq1}). 
It is not difficult to derive that $\tilde{Q}_1 = -2\sum_{(u,v)\in E}I_{u,v}^{mBIC}$ for a SD-forest $T = (V, E)$. Then we use $I_{u,v}^{mBIC}$ as edge weights and find a SD-forest (say it $M_1^{(t+1)}$ with parameter $\alpha^{(t+1)}$) with minimal $\tilde{Q}_1$ value by \citet{EdwardsAbreuLabouriau2010}'s algorithm.

To decrease $\tilde{Q}_2$ for low dimensional settings, we use stepwise procedures such as step in R. For high dimensional settings, we first get candidate selectors by applying group lasso. Specifically, for a given $h = 1, \dots, m$, we get the data set $D^{(h)}= \{y_i, x_{i,mis_i}^{(h)}, x_{i,obs_i}, i = 1, \dots, n\}$. Then we apply group lasso proposed by \citet{YuanLin2006} to obtain
$(\hat{\tau}_{\lambda}^{(h)}, \hat\beta_{\lambda}^{(h)})$ which minimize the penalized log-likelihood for the data set $D^{(h)}$
\begin{eqnarray}
\sum_{i=1}^n \log f(y_i, x_{i,mis_i}^{(h)}, x_{i,obs_i}  \mid  \tau, \beta) + \lambda \sum_{j=1}^{q+p}\sqrt{k_j}||\beta_j||_2,\nonumber
\end{eqnarray}
where $\lambda$ is a tuning parameter, $||\beta_j||_2 = (\beta_{j1}^2 + \dots + \beta_{jk_j}^2)^{1/2}$ is the $\ell_2$ norm of the parameters of dummy variables $x_j$ for $j = 1,\dots, q$, and $||\beta_j||_2 = |\beta_j|$ is the absolute value of the parameter of $x_j = z_{j-q}$ and $k_j=1$ for $j = q+1, \dots, q+p$. Group lasso encourage that whole vectors $\beta_j$ are set to zero or non-zero to select entire categorical variables. We can select a fine tuning parameter $\lambda$ and a good group lasso solution according to BIC criteria. For example, we give a series of tuning parameters $\lambda$, therefore we get a series of group lasso solutions $(\hat{\tau}_{\lambda}^{(h)}, \hat\beta_{\lambda}^{(h)})$. Furthermore, we choose one (say it, $(\hat{\tau}^{(h)}, \hat\beta^{(h)})$) with the minimum Bayesian Information criteria value over the series of group lasso solutions. Let $x_{h1}^{(h)}, x_{h2}^{(h)}, \dots, x_{hs_h}^{(h)}$ be the selected variables corresponding to the non-zero component of $(\hat{\tau}^{(h)}, \hat\beta^{(h)})$. Thus, we get a candidate variable set $\{x_{h1}^{(h)}, x_{h2}^{(h)}, \dots, x_{hs_h}^{(h)}, h = 1, \dots, m\}$. Next, we get the model (say it $M_2^{(t+1)}$ with parameter $\tau^{(t+1)}, \beta^{(t+1)}$) by the stepwise procedure such as \verb"step" in \verb"R" to select variables from the candidate variable set and minimize $\tilde{Q}_2$ in (\ref{aq2}).
In a word, we obtain $\psi^{(t+1)} = (M^{(t+1)}, \theta^{(t+1)})$ by decreasing $\tilde{Q}_1$ and $\tilde{Q}_2$ where $M^{(t+1)} = (M_1^{(t+1)}, M_2^{(t+1)})$ and $\theta^{(t+1)} = (\alpha^{(t+1)}, \tau^{(t+1)}, \beta^{(t+1)})$.

\section{Gaussian graphical model selection by the GEMS algorithm}

In this section, we consider a random vector $X = (X^{(1)},\dots,X^{(p)})$ following a multivariate normal distribution $N_p(\mu,\Sigma)$ with unknown mean $\mu$ and nonsingular covariance matrix $\Sigma$. Let $\Omega = \Sigma^{-1}$ be the inverse of the covariance matrix, then a zero entry $\Omega_{ij} =0$ if and only if $X^{(i)}$ and $X^{(j)}$ are conditionally independent given all other variables.

A Gaussian graphical model for the Gaussian random vector $X$ is represented by an undirected graph $G = (V,E)$, where the vertices $V = \{1, \dots, p\}$ represent the $p$ variables and the edges $E = (e_{ij})_{1\le i<j\le p}$ describe the conditional independence relationships among $X^{(1)}, \dots, X^{(p)}$. There is no edge between vertices $i$ and $j$ in $G$ if and only if $\Omega_{ij} = 0$, that is,
$X^{(i)}$ and $X^{(j)}$ are conditionally independent given all other variables. Thus, the Gaussian graphical model describes how these variables are mutually related.

Given complete samples $X_1,...,X_n$ of $X$, we define sample mean vector $\bar{X}$ and sample
covariance matrix $S$ by
\[
\bar{X} = n^{-1}\sum_{i=1}^n X_i \mbox{~and}\
S=n^{-1}\sum_{i=1}^n(X_i-\bar{X})(X_i-\bar{X})^\mathsf{T},
\]
respectively. The negative log-likelihood function is then expressed as
\begin{eqnarray}
l(\mu, \Omega, X) = \frac{n}{2}\ln\det{\Omega^{-1}} + \frac{n}{2}\trace\big[\Omega S\big] + \frac{n}{2}\trace\big[\Omega (\mu - \bar{X})(\mu - \bar{X})^\mathsf{T}\big].\label{loglik}
\end{eqnarray}

If the edges of the undirected graph $G$ (i.e., conditional independence relationships among $X^{(1)}, \dots, X^{(p)}$)  are known, the maximum likelihood estimation of $\mu$ and the non-zero entries of $\Omega$ can be computed by the iterative proportional scaling (IPS) procedure via minimizing the negative log-likelihood function (\ref{loglik}), see more details in \citet{Lauritzen1996}.
For high dimensional settings, we can compute maximum likelihood estimation of $\Omega$  by the improved versions of IPS (for example, IIPS, IHT and IPSP) based on junction tree or by partitioning the cliques, see \citet{XuGuoHe2011, XuGuoTang2012, XuGuoTang2015} for details.

\subsection{Model selection for Gaussian graphical model}
If the edges of the undirected graph $G$ are unknown, we wish to identify zero entries in $\Omega$. This is the problem of model selection for Gaussian graphical model. \citet{YuanLin2007} proposed to minimize the following negative $\ell_1$ norm penalized log-likelihood
\begin{eqnarray}
l(\mu = \bar X, \Omega, X) + \lambda||\Omega||_1 = \frac{n}{2}\ln\det{\Omega^{-1}} + \frac{n}{2}\trace\big[\Omega S\big] + \lambda||\Omega||_1 \label{ploglik}
\end{eqnarray}
over all positive definite matrices $\Omega$. Here, $||\Omega||_1 = \sum_{j\neq k} |\Omega_{jk}|$ is the $\ell_1$ norm penalty and $\lambda > 0$ is a tuning parameter. Different tuning parameters maybe lead to different zero entries of $\Omega$, and \citet{YuanLin2007} suggested to choose the tuning parameter that minimize the \textsc{bic} criterion. 
\citet{FriedmanHastieTibshirani2008} proposed the glasso to solve (\ref{ploglik}) using a coordinate descent procedure.

In the presence of missing data, let $o_i$ be the observed variables and $X_{io_i}$ be observed values in the $i$ th observation. $X_o = (X_{io_i})_{i=1,\dots, n}$ is the observed data. Then the negative observed log-likelihood becomes
\begin{eqnarray}
l(\mu, \Omega, X_o) = \frac{1}{2}\sum_{i=1}^{n}\big( \ln\det(\Omega^{-1})_{o_io_i} + (X_{io_i} - \mu_{o_i})^\mathsf{T}((\Omega^{-1})_{o_io_i})^{-1}(X_{io_i} - \mu_{o_i}) \big). \label{obsloglik}
\end{eqnarray}
\citet{StadlerBuhlmann2012} proposed the MissGlasso algorithm, which minimizes the following observed penalized log-likelihood
\begin{eqnarray}
l(\mu, \Omega, X_o) + \lambda||\Omega||_1, \label{pologlik}
\end{eqnarray}
by the EM algorithm for a given $\lambda$. 
In the E-step, MissGlasso computes the expected complete penalized log-likelihood by calculating the conditional expectation $E(X_{ij} \mid X_{i,o_i}, \mu^{(t)}, \Omega^{(t)}))$ and $E(X_{ij}X_{ik} \mid X_{i,o_i}, \mu^{(t)}, \Omega^{(t)}))$ for $i = 1, \dots, n$ and $j, k = 1, \dots, p$.    Here, $\mu^{(t)}$ and $\Omega^{(t)}$ are the mean and inverse covariance matrix of the current distribution, respectively. In the M-step, MissGlasso minimizes the expected complete penalized log-likelihood by the glasso in \citet{FriedmanHastieTibshirani2008}. \citet{StadlerBuhlmann2012} proved that every limit point of the sequence $\{(\mu^{(t)}, \Omega^{(t)}) | t = 0, 1, \dots\}$ generated by MissGlasso is a stationary point of the observed penalized log-likelihood in (\ref{pologlik}). However, the sequence generated by MissGlasso may not converge to the minimum points of (\ref{pologlik}).

\citet{KolarXing2012}  formed an unbiased estimator $\hat S$ of the covariance matrix from available data, and then they plugged $\hat S$ into the complete penalized log-likelihood (i.e.,  (\ref{ploglik})) via replacing $S$ by $\hat S$. This plug-in method is called the mGlasso algorithm. 
\citet{ThaiHATB2014} proposed a new Concave-Convex procedure which is however not computationally faster than the existing EM algorithm in their numerical experiments.

\subsection{The GEMS algorithm for Gaussian graphical model}\label{missggm}
For Gaussian graphical model selection under the framework of the GEMS algorithm, we adopt the Bayesian information criterion as GIC.
$$c(G, \theta_G, X) = 2 l(\mu, \Omega, X) + \ln(n)\mathrm{df}_\Omega,$$
where $G$ is a candidate graph, $\theta_G = (\mu, \Omega)$ is the parameters such that $\Omega_{ij} = 0$ if the edge between $i$ and $j$ is absent in $G$, and $\mathrm{df}_\Omega$ is the number of non-zero entries above the main diagonal of $\Omega$ (i.e., the number of edges in $G$). The observed GIC is $c(G, \theta_G, X_o) = 2 l(\mu, \Omega, X_o) + \ln(n)\mathrm{df}_\Omega$.
For the sake of simplicity, we denote $E(\cdot \mid X_{o}, \mu^{(t)}, \Omega^{(t)})$ by $E_t(\cdot)$.

\begin{algorithm}\footnotesize
\label{emsggm} 
\KwIn{An observed data matrix $X_o$ with sample size $n$ and the number of variables $p$}
\KwOut{$\hat\psi$}

Get an initial parameter $\psi^{(0)} = (G^{(0)}, \mu^{(0)}, \Omega^{(0)})$. 

\For { t= 0, 1, 2, \dots}{
     Compute conditional expectation $E_t(\bar X)$ in (\ref{ey}) and $E_t(S^*)$ in (\ref{es}) to get $Q(\psi; \psi^{(t)})$ in (\ref{Q})\;

     For small $p$, set $\mathcal G$ to be the set of all possible graphs. For large $p$, apply Glasso to (\ref{eploglik}) and  get possible sparse candidate graph set $\mathcal G$ defined in (\ref{cg})\;


    \For {each candidate graph $G \in \mathcal G$}{
         Let $\hat\mu_{G}^{(t+1)} = E_t(\bar X)$ and get $\hat\Omega_{G}^{(t+1)}= \arg\min_{\Omega_G } Q(G, \hat\mu_{G}^{(t+1)}, \Omega_G; \psi^{(t)})$ by IPS or its improved versions\;
    }
    Select $G^{(t+1)} = \arg\min_{G\in\mathcal{G}}Q(G, \hat\mu_{G}^{(t+1)}, \hat\Omega_{G}^{(t+1)}; \psi^{(t)})$ and get $\psi^{(t+1)} = (G^{(t+1)}, \hat\mu_{G^{(t+1)}}^{(t+1)}, \hat\Omega_{G^{(t+1)}}^{(t+1)})$\;

\If{$G^{(t+1)} = G^{(t)}$ or $|Q(\psi^{(t)}; \psi^{(t-1)}) - Q(\psi^{(t+1)}; \psi^{(t)})|$ is sufficient small 
     }{
    $\hat\psi = \psi^{(t+1)}$\;
    break\;}
}
\caption{GEMS for GGM}
\end{algorithm}

The GEMS algorithm for Gaussian graphical model selection is described in Algorithm \ref{emsggm}. Briefly, an initial graph with the parameter is given in Line 1. We impute the missing values by their corresponding
column means and then apply the glasso from (\ref{ploglik}) on the imputed data to obtain an initial graph. The E-step is reported in Line 3 while the MS-step is reported in Lines 4 - 7. Convergence condition is checked in Line 8.

In the E-step, the $Q$ function can be obtained as
\begin{eqnarray}
Q(\psi; \psi^{(t)}) = Q(G, \mu, \Omega; G^{(t)}, u^{(t)}, \Omega^{(t)}) = E_t(2 l(\mu, \Omega, X)) + \ln(n)\mathrm{df}_\Omega.\label{Q}
\end{eqnarray}
More details about the computation is given in Appendix.

For moderate or large $p$, it is a computational challenge to minimize the $Q$ function over all possible graphs since the number of all possible graphs is $2^{p(p-1)/2}$. In this paper, we use glasso to get candidate graph set $\mathcal G$ and choose one resulting the minimization of $Q$ function from $\mathcal G$.
Specifically, we first replace $S$ by $E_t(S^*) = n^{-1} \sum_{i=1}^n E_t\{(X_i - \mathrm{E}_t\{\bar X\})(X_i - \mathrm{E}_t\{\bar X\})^{^\mathsf{T}}\}$ in (\ref{ploglik}) and get
\begin{eqnarray}
\frac{n}{2}\ln\det{\Omega^{-1}} + \frac{n}{2}\trace\big[\Omega E_t(S^*)\big] + \lambda||\Omega||_1. \label{eploglik}
\end{eqnarray}
Then, we give an increasing positive sequence $\lambda_1, \lambda_2, \dots, \lambda_k$, and obtain $\hat\Omega_{\lambda_m}$ by using glasso to minimize (\ref{eploglik}) for each $\lambda_m$. Finally, we get the candidate graph set
\begin{eqnarray}
\mathcal G = \{ G^{(t)}, G_{\lambda_m} \mbox{ corresponding to } \hat\Omega_{\lambda_m}, m = 1, \dots, k\}.\label{cg}
\end{eqnarray}
In Lines 5 and 6 of Algorithm 1, for each candidate graph $G\in\mathcal G$, we re-estimate $\theta_G = (\mu, \Omega)$ to minimizing (\ref{Q}) satisfying that $\Omega_{ij}=0$ if and only if the edge between $i$ and $j$ is absent in $G$. Given the structure of $G\in\mathcal G$, $\log(n)\mathrm{df}_\Omega$ in (\ref{Q}) is fixed, therefore it is sufficient to minimize
\begin{eqnarray}
E_t(2 l(\mu, \Omega, X)) &=& n\ln\det{\Omega^{-1}} + n\trace\big[\Omega E_t(S^*)\big] \nonumber\\
                                                       && + n\trace\big[\Omega (\mu - E_t(\bar{X}))(\mu - E_t(\bar{X}))^\mathsf{T})\big],\label{QE}
\end{eqnarray}
which is similar to the negative log-likelihood (\ref{loglik}). In fact, $E_t(l(\mu, \Omega, X))$ in (\ref{QE}) can be obtained via replacing $\bar X$ and $S$ by $E_t(\bar X)$ and $E_t(S^*)$ in (\ref{loglik}). Therefore, we can get $\hat\mu_{G}^{(t+1)} = E_t(\bar X)$ and get $\hat\Omega_{G}^{(t+1)}= \arg\min_{\Omega} E_t(2 l(\hat\mu_{G}^{(t+1)}, \Omega, X))$ by IPS or its improved versions in Line 6. In Line 7, we select graph $G^{(t+1)}$ from $\mathcal G$ and get the parameter $\psi^{(t+1)}$. It should be noted that  it is a sufficient condition of $Q(\psi^{(t+1)}; \psi^{(t)}) \leq Q(\psi^{(t)}; \psi^{(t)})$ that $G^{(t)}$ is included in $\mathcal G$ in (\ref{cg}).

\section{Simulations}


\subsection{Simulations for variable selection in logistic regression with mixed covariates}
\begin{figure}
\centering
\epsfig{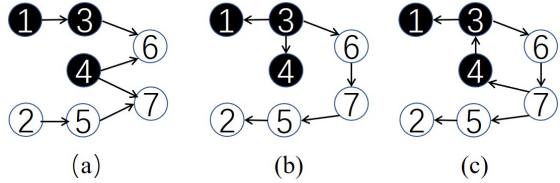}
\caption{(a) and (b) are DAGs with three categorical variables and four continuous variables, but (c) is not a DAG because it contains a cycle. Categorical variables are represented as dots and continuous variables as circles $3, 6, 7, 4, 3$. (a) is not a single parent DAG because some vertices have more than one parents, while (b) is a single parent DAG. }\label{mixedDAGs}
\end{figure}

In this subsection, we will investigate the performance of GEMS for variable selection in logistic regression.
First, we generate a directed acyclic graph (DAG) to simulate a conditional Gaussian distribution on mixed predictors. A DAG is a directed graph $\vec{G}=(V, \vec{E})$ where $V$ is a set of vertices and $\vec{E}$ is a set of directed edges (arrows) and there is no directed cycles with the arrows pointing in the same direction all the way around. Here, the vertex in $V$ represents a categorical variable or a continuous variable. If there is a directed arrow from $u$ to $v$ then $u$ is called a parent of $v$. The set of parents of $v$ is denoted by $pa(v)$. We restrict that the categorical variables do not have continuous parents in DAG $\vec{G}$. Thus, the DAG includes the single-parent DAG which is Markov equivalent to the SD-tree discussed in subsection 4.1. For example, there are two DAGs with mixed variable as its vertices in Figure \ref{mixedDAGs}. In the simulation, we use R packge pcalg to generate randomly a DAG in which each vertex connects $2$ arrows on average.

Second, we generate a conditional Gaussian distribution based on the constructed DAG. Here, its density is factorized into
\begin{eqnarray}
f(x) = \prod_{v\in V}f(x_v|x_{pa(v)}),\label{density}
\end{eqnarray}
where $x_{pa(v)}$ is the vector corresponding to the parents $pa(v)$ and $f(x_v|x_{pa(v)})$ is the conditional density of $x_v$ given the parents is equal to $x_{pa(v)}$. For example, according to Figure \ref{density} (a), we factorize
$$f(x) = f(x_1)f(x_2)f(x_3|x_1)f(x_4)f(x_5|x_2)f(x_6|x_{3}, x_{4})f(x_7|x_4, x_5).$$
Next we generate the conditional densities $f(x_v|x_{pa(v)})$ in (\ref{density}).
The number of levels of categorical variables is set to be 2 or 3 randomly. For categorical variable $W$, if it has no parent, then we draw $r_1, r_2, \dots, r_l$ from integers 2 to 8 and set the probability $P(W = i) = \frac{r_i}{\sum_{j=1}^lr_j}$ for $i = 1, \dots, l$, where $l$ is the number of levels; if $W$ has parents $W_1, W_2, \dots, W_m$, its conditional probability given $W_1=w_1, W_2=w_2, \dots, W_m=w_m$ is generated as the same as the probability when $W$ has no parent. For continuous variable $Z$, if it has no parents, then its mean and variance are drawn uniformly from the intervals $c(-1, 1)$ and $(0.5, 2)$ respectively; if $Z$ has only continuous parents $Z_1, \dots, Z_k$, then the conditional mean is $b_0 + b_1z_1 + \dots + b_kz_k$ where $b_0, \dots, b_k$ are drawn from $-1, -0.5, 0.5, 1$ with replacement and $z_1, \dots, z_k$ are values of $Z_1, \dots, Z_k$, and the conditional variance is drawn uniformly from interval $(0.5, 2)$; if $Z$ has only categorical parents $W_1, \dots, W_m$, the conditional mean and conditional variance given $W_1=w_1, W_2=w_2, \dots, W_m=w_m$ are drawn uniformly from $(-1, 1)$ and $(0.5, 2)$ respectively; if it has categorical parents $W_1, \dots, W_m$ and continuous parents $Z_1, \dots, Z_k$, given $W_1=w_1, W_2=w_2, \dots, W_m=w_m$ and $Z_1=z_1, \dots, Z_k=z_k$, the conditional mean is $b_{0w} + b_{1w}z_1 + \dots + b_{kw}z_k$ where $b_{0w}, \dots, b_{kw}$ depend on $w=(w_1, \dots, w_m)$ and they are drawn from $-1, -0.5, 0.5, 1$ with replacement, and the conditional variance depends on $w$ and it is drawn uniformly from interval $(0.5, 2)$. It is easy to know that the joint probability distribution of categorical and continuous variables is a conditional Gaussian distribution.

\begin{table}\scriptsize
\centering\caption{Results in logistics regression with mixed covariates.}\label{mixed}
\centering								
\begin{tabular}{rrrr r  rrrr rrrr r  rrrr}
\hline																								
  p, q &		&	\multicolumn{5}{c}{$10\%$}									&	&	\multicolumn{5}{c}{$20\%$}									\\
				\cline{3-7}											\cline{9-13}									
	&		&	tp	&	fp	&	tpr	&	ppv	&	mcc	&	&	tp	&	fp	&	tpr	&	ppv	&	mcc	\\
				\cline{3-7}											\cline{9-13}									
	&	imputation	&	6.23 	&	6.30 	&	0.57 	&	0.57 	&	0.53 	&	&	6.15 	&	6.18 	&	0.56 	&	0.56 	&	0.52 	\\
100	&	GEMS30	&	5.83 	&	1.66 	&	0.53 	&	0.79 	&	0.63 	&	&	5.81 	&	1.69 	&	0.53 	&	0.79 	&	0.63 	\\
	&	GEMS50	&	5.95 	&	1.74 	&	0.54 	&	0.79 	&	0.63 	&	&	5.93 	&	1.86 	&	0.54 	&	0.77 	&	0.63 	\\
	&	GEMS100	&	6.05 	&	2.01 	&	0.55 	&	0.77 	&	0.63 	&	&	6.02 	&	2.08 	&	0.55 	&	0.75 	&	0.63 	\\
\hline																								
	&		&	tp	&	fp	&	tpr	&	ppv	&	mcc	&	&	tp	&	fp	&	tpr	&	ppv	&	mcc	\\
				\cline{3-7}											\cline{9-13}									
	&	imputation	&	5.50 	&	5.92 	&	0.50 	&	0.56 	&	0.50 	&	&	5.53 	&	6.20 	&	0.51 	&	0.55 	&	0.50 	\\
200	&	GEMS30	&	5.28 	&	2.32 	&	0.48 	&	0.71 	&	0.57 	&	&	5.24 	&	2.36 	&	0.48 	&	0.71 	&	0.57 	\\
	&	GEMS50	&	5.49 	&	2.52 	&	0.50 	&	0.71 	&	0.58 	&	&	5.40 	&	2.54 	&	0.50 	&	0.69 	&	0.57 	\\
	&	GEMS100	&	5.56 	&	2.78 	&	0.51 	&	0.68 	&	0.57 	&	&	5.51 	&	2.68 	&	0.51 	&	0.69 	&	0.58 	\\
\hline																								
\end{tabular}
\end{table}								

Third, we draw randomly 3 to 5 categorical variables and 3 to 5 continuous variables to predict the binary response. The coefficients of the dummy variables and continuous variables in linear predictor of logistic regression are generated randomly from $-2, -1, 1$ and $2$.
Then we generate a dataset with 200 observations from above logistic regression model and furthermore we generate $10\%$ or $20\%$ missing values completely at random (MCAR).

For $p=q= 100, 200$, we generate 200 datasets with $10\%$ missing values and 200 datasets with $20\%$ missing values according to data-generating mechanism. The imputation method and GEMS are used to select variables for the datasets. In the imputation method, we first impute the missing value of continuous variables with its observed mean and impute the missing value of categorical variables with observed mode; then we apply group lasso to select variables in which Bayesian information criteria is used to choose the tuning parameters. In GEMS, we draw $m=30, 50, 100$ Monte Carlo samples to approximate $Q_{1i}$ and $Q_{2i}$, and the resulting GEMS is called GEMS30, GEMS50, GEMS100. To assess the performance of above methods, we evaluate the true positive rate (tpr), positive predictive value (ppv) and Matthews correlation coefficient (mcc) defined as follows
\begin{eqnarray}
tpr &=& \frac{tp}{tp+fn}, \quad  
ppv = \frac{tp}{tp+fp}, \mbox{~and} \nonumber\\   
mcc &=& \frac{tp\times tn - fp\times fn}{\{(tp+fp)(tp+fn)(tn+fp)(tn+fn)\}^{1/2}}, \label{mcc} 
\end{eqnarray}
where $tp, tn, fp$ and $fn$ are the numbers of true positives, true negatives, false positives and false negatives, respectively.
We report average tp, fp, tpr, ppv and mcc in Table \ref{mixed}. We found that tp and tpr of GEMS30, GEMS50, GEMS100 are very close to those of imputation method. Moreover, GEMS30, GEMS50, GEMS100 could reduce the false positive significantly, and could increase ppv and mcc remarkably.

\begin{table}\scriptsize
\centering
\caption{Mean and (standard deviation) for Model 1 with $p = 100, 200, 500$.}\label{msd1}		
\begin{tabular}{ll  ccccc   ccccc cc}																									
    \hline																									
		&		&			\multicolumn{3}{c}{$p=100$} 			&	&				\multicolumn{3}{c}{$p=200$} 		&	&			\multicolumn{3}{c}{$p=500$} 			\\	
								\cline{3-5}								\cline{7-9}						\cline{11-13}			
		&		&	MissGl	&	mGl	    &	GEMS	&	&	MissGl	&	mGl	    &	GEMS	&	&	MissGl	&	mGl	&	GEMS	\\	\hline																			
		&	0.1	&	1.000	&	1.000	&	0.967	&	&	1.000	&	1.000	&	0.985	&	&	1.000	&	1.000&	0.988\\	
		&		&(0.000)&(0.000)&(0.018)&	&(0.000)&(0.000)&(0.012)&	&(0.000)&(0.000)&(0.006)\\	\cline{2-13}
tpr		&	0.2	&	1.000	&	0.999	&	0.954	&	&	1.000	&	1.000	&	0.975	&	&	1.000	&	1.000&	0.978\\	
		&		&(0.001)&(0.003)&(0.033)&	&(0.000)&(0.000)&(0.018)&	&(0.000)&(0.000)&(0.011)\\	\cline{2-13}
		&	0.3	&	0.997	&	0.991	&	0.908	&	&	1.000	&	0.999	&	0.941	&	&	1.000	&	1.000&	0.968\\	
		&		&(0.006)&(0.009)&(0.040)&	&(0.001)&(0.002)&(0.022)&	&(0.000)&(0.001)&(0.011)\\	\hline
		&	0.1	&	0.199	&	0.215	&{\bf 0.662}&	&	0.220	&	0.218	&{\bf 0.694}&	&	0.238	&0.199&{\bf 0.769}\\
		&		&(0.031)&(0.031)&(0.078)&	&(0.035)&(0.019)&(0.081)&	&(0.020)&(0.030)&(0.062)\\	\cline{2-13}
ppv		&	0.2	&	0.213	&	0.255	&{\bf 0.636}&	&	0.228	&	0.256	&{\bf 0.687}&	&	0.230	&0.284&{\bf 0.754}\\			&		&(0.030)&(0.040)&(0.068)&	&(0.024)&(0.044)&(0.090)&	&(0.022)&(0.039)&(0.077)\\	\cline{2-13}
		&	0.3	&	0.238	&	0.277	&{\bf 0.616}&	&	0.235	&	0.265	&{\bf 0.674}&	&	0.227	&0.270&{\bf 0.726}\\
		&		&(0.033)&(0.049)&(0.065)&	&(0.021)&(0.056)&(0.081)&	&(0.030)&(0.080)&(0.078)\\	\hline
		&	0.1	&	0.426	&	0.445	&{\bf 0.794}&	&	0.458	&	0.457	&{\bf 0.823}&	&	0.484	&0.441&{\bf 0.870}\\
		&		&(0.036)&(0.035)&(0.046)&	&(0.039)&(0.020)&(0.046)&	&(0.021)&(0.032)&(0.035)\\	\cline{2-13}
mcc		&	0.2	&	0.443	&	0.487	&{\bf 0.772}&	&	0.468	&	0.496	&{\bf 0.814}&	&	0.476	&0.529&{\bf 0.857}\\
		&		&(0.034)&(0.043)&(0.035)&	&(0.026)&(0.046)&(0.048)&	&(0.023)&(0.039)&(0.041)\\	\cline{2-13}
		&	0.3	&	0.470	&	0.507	&{\bf 0.740}&	&	0.477	&	0.504	&{\bf 0.792}&	&	0.472	&0.511&{\bf 0.836}\\
		&		&(0.036)&(0.048)&(0.036)&	&(0.022)&(0.054)&(0.045)&	&(0.031)&(0.076)&(0.045)\\	\hline
		&	0.1	&	22.452	&	24.598	&{\bf 5.615}&	&	48.178	&	49.365	&{\bf 6.189}&	&	127.709	&122.448&{\bf 11.326}	\\	
		&		&(2.541)&(2.231)&(1.226)&	&(4.878)&(2.034)&(1.341)&	&(4.394)&(6.393)&(1.881)\\	\cline{2-13}
kl		&	0.2	&	26.644	&	31.273	&{\bf 7.077}&	&	55.518	&	61.272	&{\bf 8.465}&	&	140.692	&159.440&{\bf 16.154}	\\	
		&		&(2.833)&(3.026)&(1.973)&	&(3.355)&(4.927)&(2.150)&	&(5.673)&(8.891)&(3.327)\\	\cline{2-13}
		&	0.3	&	33.108	&	38.109	&{\bf 10.811}&	&	63.418	&	71.122	&{\bf 13.763}	&	&	157.580	&	179.451	&{\bf 	21.435}	\\	
		&		&(3.423)&(3.684)&(2.505)&	&(2.999)&(6.105)&(2.722)&	&(8.324)&(17.651)&(3.708)\\	\hline
		&	0.1	&	3.856	&	3.969	&{\bf 2.833}&	&	3.958	&	3.987	&{\bf 2.735}	&	&	4.042	&	3.993	&{\bf 	2.848}	\\	
		&		&(0.137)&(0.113)&(0.561)&	&(0.104)&(0.050)&(0.540)&	&(0.039)&(0.054)&(0.569)\\	\cline{2-13}
norm		&	0.2	&	4.071	&	4.219	&{\bf 3.076}	&	&	4.155	&	4.226	&{\bf 2.949}&	&	4.178	&	4.277	&{\bf 	3.228}	\\	
		&		&(0.113)&(0.114)&(0.670)&	&(0.068)&(0.079)&(0.583)&	&(0.041)&(0.065)&(0.458)\\	\cline{2-13}
		&	0.3	&	4.317	&	4.449	&{\bf 	3.479}	&	&	4.299	&	4.406	&{\bf 3.485}&	&	4.309	&	4.418	&{\bf 	3.422}	\\	
		&		&(0.102)&(0.098)&(0.529)&	&(0.048)&(0.074)&(0.371)&	&(0.046)&(0.085)&(0.371)\\	\hline
\end{tabular}
\end{table}

\begin{table}[t]\scriptsize
\centering
\caption{Mean and (standard deviation) for Model 2 with $p = 100, 200, 500$.}\label{msd2}
\begin{tabular}{ll  ccccc   ccccc cc}																									
    \hline																									
		&		&			\multicolumn{3}{c}{$p=100$} 			&	&				\multicolumn{3}{c}{$p=200$} 		&	&			\multicolumn{3}{c}{$p=500$} 			\\	
								\cline{3-5}								\cline{7-9}						\cline{11-13}			
		&		&	MissGl	&	mGl	&	GEMS	&	&	MissGl	&	mGl	&	GEMS	&	&	MissGl	&	mGl	&	GEMS	\\	\hline																									
		&	0.1	&	0.103	&	0.117	&{\bf 	0.190}	&	&	0.175	&	0.181	&{\bf 	0.231}	&	&	0.185	&	0.219	&{\bf 	0.248}	\\	
		&		&(0.031)&(0.025)&(0.031)&	&(0.020)&(0.032)&(0.020)&	&(0.016)&(0.011)&(0.016)\\	\cline{2-13}
tpr		&	0.2	&	0.061	&	0.078	&{\bf 	0.112	}&	&	0.118	&	0.130	&{\bf 	0.160	}&	&	0.125	&	0.127	&{\bf 	0.189	}\\	
		&		&(0.022)&(0.029)&(0.027)&	&(0.025)&(0.020)&(0.020)&	&(0.015)&(0.019)&(0.029)\\	\cline{2-13}
		&	0.3	&	0.036	&{\bf 	0.058	}&	0.048	&	&	0.067	&	0.068	&{\bf 	0.071	}&	&	0.078	&	0.097	&{\bf 	0.110	}\\	
		&		&(0.016)&(0.018)&(0.020)&	&(0.025)&(0.024)&(0.012)&	&(0.012)&(0.031)&(0.031)\\	\hline																									
		&	0.1	&{\bf 	0.791	}&	0.751	&	0.520	&	&{\bf 	0.757	}&	0.714	&	0.475	&	&{\bf 	0.816	}&	0.641	&	0.438	\\	
		&		&(0.104)&(0.084)&(0.092)&	&(0.070)&(0.135)&(0.059)&	&(0.078)&(0.056)&(0.084)\\	\cline{2-13}
ppv		&	0.2	&{\bf 	0.777	}&	0.701	&	0.592	&	&{\bf 	0.757	}&	0.712	&	0.558	&	&{\bf 	0.825	}&	0.814	&	0.505	\\	
		&		&(0.110)&(0.112)&(0.091)&	&(0.089)&(0.088)&(0.070)&	&(0.063)&(0.085)&(0.126)\\	\cline{2-13}
		&	0.3	&{\bf 	0.722	}&	0.601	&	0.672	&	&{\bf 	0.758	}&	0.747	&	0.713	&	&{\bf 	0.780	}&	0.673	&	0.611\\	
		&		&(0.125)&(0.104)&(0.117)&	&(0.106)&(0.115)&(0.068)&	&(0.058)&(0.142)&(0.141)\\	\hline																									
		&	0.1	&	0.262	&	0.275	&{\bf 	0.277	}&	&{\bf 	0.352	}&	0.342	&	0.311&	&{\bf 	0.383	}&	0.369	&	0.318\\	
		&		&(0.031)&(0.027)&(0.021)&	&(0.015)&(0.013)&(0.015)&	&(0.012)&(0.010)&(0.028)\\	\cline{2-13}
mcc		&	0.2	&	0.200	&	0.212	&{\bf 	0.230	}&	&	0.286	&{\bf 	0.291	}&	0.283	&	&{\bf 	0.317	}&	0.316	&	0.297\\	
		&		&(0.033)&(0.035)&(0.028)&	&(0.022)&(0.016)&(0.015)&	&(0.013)&(0.011)&(0.018)\\	\cline{2-13}
		&	0.3	&	0.145	&{\bf 	0.165	}&	0.159&	&	0.210	&	0.212	&{\bf 	0.216	}&	&	0.243	&	0.243	&{\bf 	0.246	}\\	
		&		&(0.029)&(0.027)&(0.032)&	&(0.036)&(0.021)&(0.016)&	&(0.013)&(0.021)&(0.013)\\	\hline																									
		&	0.1	&	17.678	&	17.818	&{\bf 	16.663	}&	&	31.483	&	31.825	&{\bf 	28.653}	&	&	77.739	&	73.890	&{\bf 	69.638}	\\	
		&		&(0.966)&(0.686)&(1.856)&	&(0.983)&(2.024)&(2.014)&	&(2.538)&(1.767)&(11.164)\\	\cline{2-13}
kl		&	0.2	&	18.351	&	19.157	&{\bf 	17.469	}&	&	33.633	&	35.048	&{\bf 	30.136	}&	&	83.376	&	88.213	&{\bf 	73.377	}\\	
		&		&(0.696)&(1.024)&(1.425)&	&(1.344)&(1.141)&(1.905)&	&(1.788)&(2.786)&(10.123)\\	\cline{2-13}
		&	0.3	&	18.971	&	20.461	&{\bf 	17.698	}&	&	35.551	&	39.532	&{\bf 	31.242	}&	&	86.464	&	92.871	&{\bf 	73.965	}\\	
		&		&(0.736)&(0.822)&(1.214)&	&(1.272)&(1.776)&(0.968)&	&(1.406)&(4.756)&(4.263)\\	\hline																									
		&	0.1	&	2.225	&	2.228	&{\bf 	1.678}	&	&	2.180	&	2.185	&{\bf 	1.619}	&	&	2.181	&	2.160	&{\bf 	1.596}\\	
		&		&(0.028)&(0.018)&(0.084)&	&(0.015)&(0.029)&(0.050)&	&(0.014)&(0.011)&(0.043)\\	\cline{2-13}
norm		&	0.2	&	2.242	&	2.258	&{\bf 	1.834}	&	&	2.210	&	2.228	&{\bf 	1.774}	&	&	2.212	&	2.236	&{\bf 	1.749}	\\	
		&		&(0.022)&(0.027)&(0.081)&	&(0.019)&(0.016)&(0.065)&	&(0.010)&(0.016)&(0.079)\\	\cline{2-13}
		&	0.3	&	2.256	&	2.285	&{\bf 	1.978}	&	&	2.234	&	2.278	&{\bf 	1.965}	&	&	2.228	&	2.258	&{\bf 	1.926}	\\	
		&		&(0.019)&(0.016)&(0.054)&	&(0.018)&(0.020)&(0.036)&	&(0.008)&(0.021)&(0.064)\\	\hline
\end{tabular}
\end{table}

\begin{table}[t]\scriptsize
\caption{Mean and (standard deviation) for Model 3 with $p = 100, 200, 500$.}	\label{msd3}																							\begin{tabular}{ll  ccccc   ccccc cc}																									
    \hline																									
		&		&			\multicolumn{3}{c}{$p=100$} 			&	&				\multicolumn{3}{c}{$p=200$} 		&	&			\multicolumn{3}{c}{$p=500$} 			\\	
								\cline{3-5}								\cline{7-9}						\cline{11-13}			
		&		&	MissGl	&	mGl	&	GEMS	&	&	MissGl	&	mGl	&	GEMS	&	&	MissGl	&	mGl	&	GEMS	\\	\hline
		&	0.1	&	0.415	&{\bf 	0.439	}&	0.393	&	&	0.676	&{\bf 	0.678	}&	0.584	&	&{\bf 	0.726	}&	0.730	&	0.665\\	
		&		&(0.083)&(0.076)&(0.061)&	&(0.060)&(0.061)&(0.045)&	&(0.050)&(0.054)&(0.036)\\	\cline{2-13}
tpr		&	0.2	&	0.277	&{\bf 	0.323	}&	0.196	&	&	0.490	&{\bf 	0.515	}&	0.382	&	&	0.564	&{\bf 	0.577	}&	0.425\\	
		&		&(0.052)&(0.057)&(0.056)&	&(0.069)&(0.068)&(0.051)&	&(0.059)&(0.055)&(0.048)\\	\cline{2-13}
		&	0.3	&	0.197	&{\bf 	0.242	}&	0.095	&	&	0.294	&{\bf 	0.353	}&	0.194	&	&	0.374	&{\bf 	0.422	}&	0.214	\\	
		&		&(0.045)&(0.058)&(0.025)&	&(0.049)&(0.055)&(0.039)&	&(0.059)&(0.057)&(0.031)\\	\hline
		&	0.1	&	0.501	&	0.508	&{\bf 	0.528}	&	&	0.461	&	0.466	&{\bf 	0.534}	&	&	0.404	&	0.414	&{\bf 	0.465}	\\	
		&		&(0.051)&(0.056)&(0.042)&	&(0.041)&(0.038)&(0.038)&	&(0.032)&(0.043)&(0.032)\\	\cline{2-13}
ppv		&	0.2	&	0.470	&	0.482	&{\bf 	0.522}	&	&	0.476	&	0.495	&{\bf 	0.565}	&	&	0.399	&	0.434	&{\bf 	0.506}	\\	
		&		&(0.042)&(0.051)&(0.048)&	&(0.041)&(0.046)&(0.036)&	&(0.028)&(0.038)&(0.029)\\	\cline{2-13}
		&	0.3	&	0.436	&	0.464	&{\bf 	0.530}	&	&	0.479	&	0.500	&{\bf 	0.585}	&	&	0.375	&	0.427	&{\bf 	0.537}	\\	
		&		&(0.058)&(0.064)&(0.080)&	&(0.047)&(0.041)&(0.051)&	&(0.028)&(0.038)&(0.038)\\	\hline
		&	0.1	&	0.427	&{\bf 	0.443	}&	0.430	&	&	0.542	&{\bf 	0.546	}&	0.545	&	&	0.535	&	0.542	&{\bf 	0.550	}\\	
		&		&(0.041)&(0.032)&(0.033)&	&(0.017)&(0.016)&(0.021)&	&(0.011)&(0.013)&(0.016)\\	\cline{2-13}
mcc		&	0.2	&	0.334	&{\bf 	0.366	}&	0.297	&	&	0.466	&{\bf 	0.489	}&	0.451	&	&	0.467	&{\bf 	0.493	}&	0.458	\\	
		&		&(0.037)&(0.029)&(0.049)&	&(0.026)&(0.023)&(0.030)&	&(0.014)&(0.012)&(0.022)\\	\cline{2-13}
		&	0.3	&	0.267	&{\bf 	0.307	}&	0.207	&	&	0.360	&{\bf 	0.405	}&	0.325	&	&	0.366	&{\bf 	0.416	}&	0.334\\	
		&		&(0.037)&(0.039)&(0.033)&	&(0.028)&(0.028)&(0.038)&	&(0.022)&(0.016)&(0.025)\\	\hline
		&	0.1	&	21.930	&	22.060	&{\bf 	16.464}	&	&	44.446	&	45.515	&{\bf 	28.914}	&	&	101.696	&	104.678	&{\bf 	58.426}	\\	
		&		&(1.955)&(1.955)&(0.803)&	&(3.666)&(3.516)&(1.424)&	&(6.994)&(8.104)&(2.296)\\	\cline{2-13}
kl		&	0.2	&	24.238	&	24.783	&{\bf 	19.688	}&	&	53.068	&	54.891	&{\bf 	38.335	}&	&	118.433	&	125.411	&{\bf 	81.999}	\\	
		&		&(1.210)&(1.562)&(0.992)&	&(3.638)&(3.984)&(2.301)&	&(7.576)&(7.780)&(4.889)\\	\cline{2-13}
		&	0.3	&	25.803	&	27.410	&{\bf 	21.981	}&	&	61.811	&	64.090	&{\bf 	49.147	}&	&	138.198	&	146.450	&{\bf 	110.894	}\\	
		&		&(1.275)&(2.025)&(0.793)&	&(3.105)&(3.397)&(2.203)&	&(8.067)&(8.425)&(4.666)\\	\hline
		&	0.1	&	3.986	&	3.974	&{\bf 	2.566}	&	&	3.919	&	3.945	&{\bf 	2.276}	&	&	3.574	&	3.604	&{\bf 	2.035}	\\	
		&		&(0.112)&(0.119)&(0.157)&	&(0.106)&(0.102)&(0.138)&	&(0.094)&(0.100)&(0.106)\\	\cline{2-13}
norm		&	0.2	&	4.142	&	4.133	&{\bf 	3.161}	&	&	4.166	&	4.191	&{\bf 	2.803}	&	&	3.797	&	3.852	&{\bf 	2.579}	\\	
		&		&(0.065)&(0.080)&(0.231)&	&(0.093)&(0.099)&(0.172)&	&(0.089)&(0.084)&(0.141)\\	\cline{2-13}
		&	0.3	&	4.222	&	4.240	&{\bf 	3.522}	&	&	4.392	&	4.401	&{\bf 	3.395}	&	&	4.041	&	4.076	&{\bf 	3.201}	\\	
		&		&(0.057)&(0.082)&(0.141)&	&(0.071)&(0.071)&(0.170)&	&(0.082)&(0.080)&(0.102)\\	\hline
\end{tabular}									
																									
\end{table}

\subsection{Simulations for Gaussian graphical model selection}
In this subsection, we compare our method based on GEMS with the MissGlasso (abbreviated as MissGl) and mGlasso (abbreviated as mGl) methods. 
For MissGlasso and mGlasso, the tuning parameter $\lambda$ is chosen to minimize the \textsc{bic} criteria based on the observed log-likelihood (\ref{obsloglik}). 
The following models are considered:
\begin{description}
    \item[Model 1.] An autoregressive model of order 1 with $(\Omega^{-1})_{jk} = 0.7^{|j-k|}$.

    \item[Model 2.] An autoregressive model of order 4 with $\Omega_{jk} = \mathbb I_{\{|j-k|=0\}} + 0.4\mathbb I_{\{|j-k|=1\}}  + 0.2\mathbb I_{\{|j-k|=2\}}  + 0.2\mathbb I_{\{|j-k|=3\}}  + 0.1\mathbb I_{\{|j-k|=4\}}$, where $\mathbb I$ represents the indictor function.

    \item[Model 3.] A model with $\Omega = \mathrm B + \delta\mathrm I$, where each off-diagonal entry in $\mathrm B$ is generated independently and equals 0.5 with probability $\alpha = 5/p$ or 0 with probability $1 - \alpha$. Diagonal entries of $\mathrm B$ is zero and $\delta$ is chosen so that the condition number of $\Omega$ is $p$. Note that $\alpha = 5/p$ will result in a sparse model with average 5 neighbours for each variable.
\end{description}

%

We first show the experiment results under the missing completely at random mechanism. The number of variables and sample size are set as $(p, n) = (100, 100), (200, 150), (500, 200)$ for each model. For each simulated data set, $10\%, 20\%$ and $30\%$ of the entries are removed completely at random.  To compare these methods, we evaluate the true positive rate (tpr), positive predictive value (ppv) and Matthews correlation coefficient (mcc) defined in \ref{mcc}.
The Kullback-Leibler divergence (denoted by $kl$) between the estimated distributions obtained by the above methods and the true distribution, and the difference between the estimated $\hat\Omega$ and the true $\Omega$ based on the $\Vert \hat\Omega - \Omega \Vert_2$ 2-norm (denoted by $norm$) are also evaluated.

The means and standard deviations of the above measures based on Models 1 to 3 over 50 independent runs for each setting are reported in Tables \ref{msd1} to \ref{msd3}, respectively. Except for $tpr$, it can be seen that GEMS generally outperforms MissGlasso and mGlasso for all measures for Model 1 (see, Table \ref{msd1}). Moreover, the smallest value of tprs of GEMS for Model 1 is $0.908$ when $(p, n) = (100, 100)$ and missing rate is equal to $0.3$. From Tables 2 and 3, GEMS performs better than MissGlasso and mGlasso. 
When mcc is used as a measure, no substantial difference is observed among methods in all cases for Model 2; but, mGlasso performs better for most settings except for the case of $(p, n) = (500, 200)$ and missing rate being $0.1$ for Model 3.

\begin{table}[t]\scriptsize
\centering\caption{Average CPU time in seconds. }\label{cputime}	
\begin{tabular}{ll  rrrr}													
\hline													
			&		&	MissGl	&	mGl	&	GEMS		\\
	\hline												
			&	$p=100$	&	5.3	&	1.5	&	2.7	\\
		Model 1	&	$p=200$	&	30.7	&	10.8	&	19.7		\\
			&	$p=500$	&	464.7	&	182.5	&	314.2	\\
	\hline												
			&	$p=100$	&	2.3	&	1.1	&	2.4	\\
		Model 2	&	$p=200$	&	13.8	&	7.6	&	18.9	\\
			&	$p=500$	&	202.1	&	123.0	&	280.3   \\
	\hline												
			&	$p=100$	&	4.1	&	1.4	&	2.6	\\
		Model 3	&	$p=200$	&	25.6	&	11.0	&	22.0	\\
			&	$p=500$	&	456.7	&	176.9	&	343.1	\\
\hline													
\end{tabular}												
\end{table}

Table \ref{cputime} shows the average CPU time (in second) for all methods. Obviously, GEMS and MissGlasso run slower than mGlasso as they are iterative methods. For Models 1 and 3, GEMS needs shorter CPU time than MissGlasso. For Model 2, MissGlasso requires less CPU time. 


\begin{table}[t]\scriptsize
\centering\caption{Mean (and standard deviation) for three missing mechanisms.}\label{msdMechanism123}
\begin{tabular}{lc  ccccc   ccccc ccccc}
    \hline																									
		&	$\pi$	&			\multicolumn{3}{c}{mechanism 1} 			&	&				\multicolumn{3}{c}{mechanism 2} 		&	&			\multicolumn{3}{c}{mechanism 3} 			\\	
								\cline{3-5}								\cline{7-9}						\cline{11-13}			
		&		&	MissGl	&	mGl	&	GEMS	&	&	MissGl	&	mGl	&	GEMS	&	&	MissGl	&	mGl	&	GEMS	\\	\hline
		&	$0.25$	&	1.000 	&	1.000 	&	0.994 	&	&	1.000 	&	1.000 	&	0.994 	&	&	1.000 	&	1.000 	&	0.951 	\\	
		&		&(0.000)&(0.000)&(0.016)&	&(0.000)&(0.000)&(0.016)&	&(0.000)&(0.000)&(0.049)\\	\cline{2-13}
tpr		&	$0.50$	&	0.998 	&	1.000 	&	0.974 	&	&	0.981 	&	0.998 	&	0.879 	&	&	0.821 	&	0.930 	&	0.566 	\\	
		&		&(0.010)&(0.000)&(0.032)&	&(0.039)&(0.010)&(0.062)&	&(0.105)&(0.069)&(0.079)\\	\cline{2-13}
		&	$0.75$	&	0.874 	&	0.978 	&	0.685 	&	&	0.676 	&	0.937 	&	0.521 	&	&	0.528 	&	0.714 	&	0.500 	\\	
		&		&(0.095)&(0.034)&(0.102)&	&(0.092)&(0.060)&(0.030)&	&(0.041)&(0.084)&(0.000)\\	\hline
		&	$0.25$	&	0.257 	&	0.233 	&	{\bf 0.811}	&	&	0.255 	&	0.290 	&	{\bf 0.890}	&	&	0.266 	&	0.331 	&	{\bf 0.811}	\\	
		&		&(0.070)&(0.038)&(0.102)&	&(0.067)&(0.078)&(0.099)&	&(0.064)&(0.083)&(0.109)\\	\cline{2-13}
ppv		&	$0.50$	&	0.282 	&	0.237 	&	{\bf 0.828}	&	&	0.331 	&	0.317 	&	{\bf 0.831}	&	&	0.281 	&	0.361 	&	{\bf 0.836}	\\	
		&		&(0.067)&(0.049)&(0.095)&	&(0.090)&(0.073)&(0.091)&	&(0.051)&(0.097)&(0.121)\\	\cline{2-13}
		&	$0.75$	&	0.316 	&	0.236 	&	{\bf 0.846}	&	&	0.317 	&	0.283 	&	{\bf 0.788}	&	&	0.185 	&	0.336 	&	{\bf 0.833}	\\	
		&		&(0.077)&(0.045)&(0.133)&	&(0.064)&(0.070)&(0.149)&	&(0.053)&(0.088)&(0.129)\\	\hline
		&	$0.25$	&	0.464 	&	0.440 	&{\bf	0.890} 	&	&	0.463 	&	0.500 	&{\bf	0.936} 	&	&	0.476 	&	0.541 	&{\bf	0.869} 	\\	
		&		&(0.075)&(0.044)&(0.061)&	&(0.073)&(0.077)&(0.057)&	&(0.068)&(0.079)&(0.064)\\	\cline{2-13}
mcc		&	$0.50$	&	0.491 	&	0.444 	&{\bf	0.891} 	&	&	0.534 	&	0.528 	&{\bf	0.845} 	&	&	0.439 	&	0.544 	&{\bf	0.673} 	\\	
		&		&(0.070)&(0.056)&(0.055)&	&(0.077)&(0.068)&(0.057)&	&(0.052)&(0.082)&(0.076)\\	\cline{2-13}
		&	$0.75$	&	0.484 	&	0.437 	&{\bf	0.746} 	&	&	0.423 	&	0.474 	&{\bf	0.623} 	&	&	0.254 	&	0.450 	&{\bf	0.631} 	\\	
		&		&(0.057)&(0.053)&(0.088)&	&(0.055)&(0.068)&(0.063)&	&(0.047)&(0.076)&(0.055)\\	\hline
		&	$0.25$	&	3.479 	&	3.579 	&{\bf	0.862} 	&	&	3.712 	&	4.944 	&{\bf	1.030} 	&	&	5.095 	&	6.327 	&{\bf	4.490 	}\\	
		&		&(0.718)&(0.514)&(0.321)&	&(0.827)&(0.688)&(0.472)&	&(0.646)&(0.612)&(1.005)\\	\cline{2-13}
kl		&	$0.50$	&	4.680 	&	4.934 	&{\bf	1.369} 	&	&	5.688 	&	6.808 	&{\bf	3.770 	}&	&{\bf	9.169 	}&	10.458 	&	16.384 	\\	
		&		&(0.813)&(0.673)&(0.539)&	&(0.978)&(0.614)&(1.047)&	&(0.820)&(0.889)&(2.325)\\	\cline{2-13}
		&	$0.75$	&	7.285 	&	8.968 	&{\bf	6.485} 	&	&{\bf	8.313 	}&	10.893 	&	10.242 	&	&{\bf	10.214 	}&	16.501 	&	32.557 	\\	
		&		&(1.120)&(1.469)&(1.767)&	&(0.627)&(1.742)&(1.305)&	&(0.924)&(2.879)&(5.202)\\	\hline
		&	$0.25$	&	2.545 	&	2.608 	&{\bf	1.618} 	&	&	2.528 	&	2.731 	&{\bf	1.457} 	&	&	2.514 	&	2.734 	&{\bf	2.031} 	\\	
		&		&(0.207)&(0.153)&(0.688)&	&(0.191)&(0.150)&(0.382)&	&(0.161)&(0.144)&(0.409)\\	\cline{2-13}
norm		&	$0.50$	&	2.750 	&	2.843 	&{\bf	1.711} 	&	&	2.822 	&	2.991 	&{\bf	2.107} 	&	&{\bf	2.706 	}&	2.831 	&	3.051\\	
		&		&(0.153)&(0.144)&(0.414)&	&(0.152)&(0.108)&(0.254)&	&(0.136)&(0.137)&(0.666)\\	\cline{2-13}
		&	$0.75$	&	3.018 	&	3.391 	&{\bf	2.300} 	&	&	3.079 	&	3.576 	&{\bf	2.359} 	&	&{\bf	2.769 	}&	3.688 	&	5.677 \\	
		&		&(0.153)&(0.595)&(0.176)&	&(0.106)&(0.753)&(0.368)&	&(0.156)&(1.764)&(1.477)\\	\hline
\end{tabular}																							

\end{table}

In the next experiment, we will show the performance of all method when the missing values are generated at random. 
The following model will be considered
\begin{description}
    \item[Model 4.] A Gaussian graphical model with $p = 30$ and a block-diagonal covariance matrix $\Sigma = \mathrm{diag}(B, B, \dots, B)$ where $B\in R^{3\times 3}$ and $B_{jk} = 0.7^{|j - k|}$.
\end{description}
It is noted that this model was also considered in \citet{StadlerBuhlmann2012} and \citet{KolarXing2012}. Specifically, we generate data set with sample size $n = 100$ and delete values from the data set according to the following missing data mechanisms:
\begin{enumerate}
    \item For all $b = 1, \dots, 10$ and $i = 1, \dots, n$, $X_{i, 3*b}$ is missing if $R_{i,j} = 0$ where $R_{i,j}$ follows a Bernoulli distribution with probability $\pi$.

    \item For all $b = 1, \dots, 10$ and $i = 1, \dots, n$, $X_{i, 3*b}$ is missing if $X_{i, 3*b-2} < T$

    \item For all $b = 1, \dots, 10$ and $i = 1, \dots, n$, $X_{i, 3*b}$ is missing if $X_{i, 3*b} < T$
\end{enumerate}
It is obviously that the threshold value (i.e., $T$) determines the percentage of missing values. We consider three settings: (a) $\pi = 0.25$ with $T = \Phi^{-1}(0.25)$, (b)  $\pi = 0.50$ with $T = \Phi^{-1}(0.50)$, and (c)  $\pi = 0.75$ with $T = \Phi^{-1}(0.75)$, where $\Phi(\cdot)$ is the standard normal cumulative distribution function. Mechanisms 1, 2 and 3 are respectively the missing completely at random (MCAR), missing at random (MAR) and not missing at random (NMAR). We report the means and standard deviation of the above measures over 50 independent runs for each missing mechanism in Table \ref{msdMechanism123}. From Table \ref{msdMechanism123},  if mcc is used as a measure, GEMS outperforms MissGlasso and mGlasso for all three mechanisms. If the Kullback-Leibler divergence and 2-norm are used as a measure, GEMS performs the best at Mechanisms 1 and 2. However, MissGlasso works better than GEMS and mGlasso for Mechanism 3.

\section{Real data analysis}

\subsection{Horse colic data}
In this subsection we will analyze horse colic data set in UCI Machine Learning Repository.
It is available at \url{http://archive.ics.uci.edu/ml/datasets/Horse+Colic}. The training set consists 299 instances with 28 attributes. We delete the 25th to 28th attributes representing type of lesion since they are a little bit confusing. We delete the horses' hospital Number. We consider a binary response $y$ defined as $y=0$ if the horse lived and $y=1$ otherwise. The rest 22 attributes includes seven continuous variables and fifteen categorical variables. The data contains many missing values. Nineteen attributes of the rest 22 attributes have missing values. In all 299 instances, 293 instances have missing values. We apply the imputation method combined with group lasso to select variables, in which we choose the tuning parameter by BIC criteria. The imputation method selects three continuous variables: pulse, packed cell volume, total protein, and three categorical variables: temp of extremities with 4 levels, pain with 5 levels, surgical lesion with 2 levels. By contrast, GEMS30 and GEMS200 select two continuous variables: pulse, packed cell volume, and one categorical variable: surgical lesion.

\subsection{Prostate cancer data}
In this subsection  we will analyze prostate cancer data (GEO GDS3289). It is available at \url{https://www.ncbi.nlm.nih.gov/geo/}. The dataset includes 34 benign epithelium samples and 70 non-benign samples. We consider a binary response $(y)$ defined as $y=1$ if it is a benign sample and $y=0$ if otherwise. We choose the first 625 biomarkers in platform ``Hs6-1-1-1" to ``Hs6-1-25-25" as covariates. In the dataset, 82 biomarkers contain no missing values, while 543 biomarkers contain missing values in 104 samples. The biomarker ``MLL" in platform
``Hs6-1-3-1" has only one observed value and the biomarker ``IMAGE:366953" in ``Hs6-1-1-18" has no observed values, therefore we remove ``MLL" and ``IMAGE:366953". Moreover, all of the 104 samples contain missing values and the missing value percentages range from $6.4\%$ to $47.2\%$. 	
\begin{table}[t]\scriptsize
\centering\caption{Results on variable selection for the prostate cancer data.}\label{cancer}
\centering								
\begin{tabular}{rl}
\hline				
method & selected biomarkers\\		
\hline																		
imputation 	& MEF2A, RCC1, IMAGE:133130, IMAGE:196837, IMAGE:200418, IMAGE:295599, \\
    & IMAGE:206867, GCLM, IMAGE:296033, IMAGE:430233, ZC3H12C, DOK1,  RGS7BP,\\
 &  IMAGE:40728, AGRN, ZNF598, EFCAB6, IMAGE:470914, KIF9, CAPRIN1,\\
 & IMAGE:773430, IMAGE:30959, TRIM5, SYNJ1, PHACTR2, SPAG11A, APBB2, \\
 & PRSS8, ZNF124, STMN1, ACOX3, CYP3A5\\
\hline
GEMS30	&	RCC1, IMAGE:133130, KIF2C, DOK1	\\
\hline
GEMS200	&	RCC1, IMAGE:133130, KIF2C, DOK1	\\
\hline																								
\end{tabular}
\end{table}								

We suppose the values of biomarkers follow multivariate normal distribution and we apply GEMS and imputation method combined with group lasso to select variables. In GEMS we use $m=30, 200$ Monte Carlo samples to approximate $Q_{1i}$ and $Q_{2i}$. The imputation method selects 32 biomarkers, which seems to be consistent with that the imputation method select a large number of false positive variables. By comparison, GEMS30 and GEMS200 selects the same four biomarkers.

\subsection{Yeast cell expression data}
\citet{{GaschSKCESBB2000}} used an yeast cell expression data to explore genomic expression patterns in the yeast Saccharomyces cerevisiae.
This data set contains $p$ = $6152$ known or predicted yeast genes and the sample size is $n$ = $173$. It is available at \url{http://genome-www.stanford.edu/yeast_stress/}. The whole data set has about 3.01\% missing values. There is no complete data record, 129 records with more than 50 missing values, and the maximum number of missing values is 1384.  Only 755 genes have no missing values, and 0.53\% genes have at least 3 missing values. For illustration purpose, we will use our proposed GEMS to study the regulatory relationships between the 6152 yeast genes by Gaussian graphical model.

\begin{figure}
\centering
\epsfig{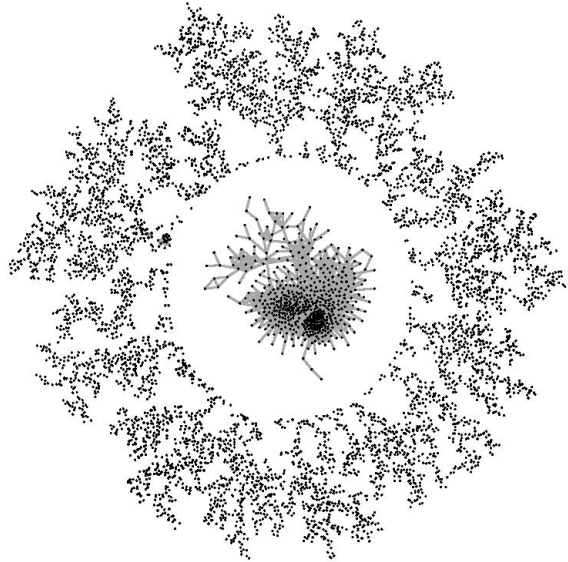}
\caption{The graphical structure obtained by GEMS for the yeast data.}\label{yeast}
\end{figure}

The extended Bayesian information criterion proposed by \citet{ChenChen2008} will be used as the generalized information criteria since \citet{FoygelDrton2010} established the consistency of the extended \textsc{bic} for Gaussian graphical models under some conditions. Specifically, the extended \textsc{bic} is given by
\begin{eqnarray}\label{ebic}
2 l(G, \mu_G, \Omega_G, X_o) + \mathrm{df}_{\Omega_G}\log(n) + 4\mathrm{df}_{\Omega_G}\gamma\log(p),
\end{eqnarray}
where $l(G, \mu_G, \Omega_G, X_o)$ is minus twice the observed log-likelihood and $\mathrm{df}_G$ is the number of free parameters of $\Omega_G$. In this example, we choose $\gamma = 0.25$ and adopt the initial $\Sigma$ to be the diagonal matrix with the main diagonal entries being the variances of all genes. GEMS ran 15564.93 seconds to obtain the graphical structure with 8703 edges shown in Figure \ref{yeast}. This structure includes 5526 connected components and the maximal connected component has 552 genes. In this structure, 5482 genes have no neighbor and 36 genes have more than 100 neighbors.

We compare our resultant graphical structure with YeastNet proposed by \citet{KimSKKHSL2013}. In particular, YeastNet covers up to 5818 genes which are wired by 362512 functional links (edges). It is found that our graphical structure and YeastNet share $5493$ common genes. We compare sub-graphical structure and sub-YeastNet with 5493 genes. Besides, while our sub-graphical structure includes 3697 edges which are also identified by sub-YeastNet, our sub-graphical structure finds an extra 3590 edges which are not included in the sub-YeastNet. Interestingly, GEMS concludes that the top three strongly correlated gene pairs are (YHR215W, YAR071W), (YDR343C, YDR342C) and (YHR055C, YHR053C) with  the  estimated partial correlations being $-0.97$, $-0.97$ and $-0.98$, respectively. In YeastNet, these three gene pairs are assigned with very high log likelihood scores being $5.00, 5.47$ and $4.36$, respectively. 

\section{Conclusion}

In this paper, we proposed a generalized EMS (GEMS) algorithm, which includes the EMS algorithm as a special case. We prove the numerical convergence of the GEMS algorithm. Furthermore, we prove in Corollary \ref{emsC1} that all limit points of the EMS algorithm satisfy a necessary condition of the minimum points of the observed GIC under relatively weak conditions, which are much more useful in practice. We apply the GEMS algorithm for Gaussian graphical model selection and generalized linear model selection with missing data. For generalized linear model with both categorical predictors and continuous predictors, we model the predictors by SD-tree or SD-forest for computing $Q$ function in high dimensional settings. The simulation studies further confirm that GEMS outperforms the existing competitors.

\section*{Acknowledgements}
We thank the authors of \citet{StadlerBuhlmann2012} and \citet{AbreuEdwardsLabouriau2010}
for sharing their R codes and R packages.
PF Xu and N Shan's work were supported by the National Natural Science Foundation of China (Grant No. 11871013). ML Tang's work was supported by the grants from the Research Grant Council of the Hong Kong Special Administrative Region (Project Nos. UGC/FDS14/P01/17, UGC/FDS14/P02/18). 



\appendix
\section{Proofs of our results}

To prove the convergence of the GEMS algorithm, we need to define a point-to-set mapping and briefly introduce the Global Convergence Theorem. According to \citet{LuenbergerYe2008}, an iterative algorithm is a mapping $\mathbf A$ defined on a space $X$ that assigns to every point $x\in X$ a subset of $X$. Here, $\mathbf A$ is a point-to-set mapping of $X$ which generalizes a point-to-point mapping of $X$.
A point-to-set mapping $\mathbf A$ is said to be closed at $x\in X$ if $x_k \rightarrow x$ for $x_k\in X$ and $y_k \rightarrow y$ for $y_k\in \mathbf A(x_k)$ imply $y\in \mathbf A(x)$.
If the set $\mathbf A(x)$ consists of a single point and $\mathbf A$ is continuous, then $\mathbf A$ is closed.

\begin{theorem}[Global Convergence Theorem]\label{appthm1}
Let $\mathbf{A}$ be an algorithm on $X$. Suppose that, given $x_0$, the sequence $\{x_k\}^\infty_{k=0}$ is generated and satisfies $x_{k+1} \in A(x_k)$. Let the solution set $\Gamma\subset X$ be given. If $\mathrm{(\romannumeral 1)}$ all points $x_k$ are contained in a compact set $S\subset  X$; $\mathrm{(\romannumeral 2)}$ there is a continuous function $Z$ on $X$ such that $(a)$ if $x \not\in \Gamma$, then $Z(y) < Z(x)$ for all $y\in \mathbf{A}(x)$, and $(b)$ if $x\in\Gamma$, then $Z(y) \leq Z(x)$ for all $y\in \mathbf{A}(x)$; and $\mathrm{(\romannumeral 3)}$ the mapping $\mathbf A$ is closed at points outside $\Gamma$, then the limit of any convergent subsequence of $\{x_k\}$ is a solution in $\Gamma$.
\end{theorem}

Here, the function $Z$ could be the objective function to be minimized. It should be noticed that $Z$ is continuous if $x_k \rightarrow x$ implies $Z(x_k)\rightarrow Z(x)$. The solution set $\Gamma$ could be the set of minimum points of the objective function or the set of points satisfying the necessary condition of minimum points (e.g., stationary points of the objective function). For more details about the Global Convergence Theorem, one can refer to Section 7.7 of \citet{LuenbergerYe2008}.
It should be noticed that the point-to-set mapping in GEMS is $\mathbf A(\psi) = \{\tilde\psi | Q(\tilde\psi; \psi) \leq Q(\psi; \psi)\}$ while the corresponding  mapping in EMS is $\mathbf A(\psi) = \arg\min_{\psi'} Q(\psi'; \psi)$ for any $\psi\in\Psi$. In both algorithms, the function $Z$ is taken to be the observed GIC $g(\psi, Y_o)$.

Next, we give proofs of our results.

{\it Proof of Theorem \ref{neceCond}}. For any $\psi^*\in \Psi_0$ and $\psi\in \Psi$, we have,  by (\ref{gexp}), $Q(\psi^*; \psi^*) = g(\psi^*, Y_o) + H(\psi^*; \psi^*)$ and $Q(\psi; \psi^*) = g(\psi, Y_o) + H(\psi; \psi^*)$.
According to Condition 3 and $g(\psi^*, Y_o)  \leq g(\psi, Y_o)$, we have $Q(\psi^*; \psi^*) \leq Q(\psi; \psi^*)$; i.e., $\psi^*\in \Psi_1$. $\hfill{} \Box$

{\it Proof of Theorem \ref{non-increasing}}.
By Conditoin 3 and equation (\ref{gexp}), we have  for any $\tilde\psi\in \mathbf A(\psi)$
\begin{eqnarray}
g(\psi, Y_o) = Q(\psi; \psi) - H(\psi; \psi) \geq Q(\tilde\psi; \psi) - H(\tilde\psi; \psi) = g(\tilde\psi, Y_o),\nonumber
\end{eqnarray}
where the equality holds if and only if $Q(\psi; \psi) = Q(\tilde\psi; \psi)$ and $H(\psi; \psi) = H(\tilde\psi; \psi)$. $\hfill{} \Box$

{\it Proof of Theorem \ref{gemsC}}.
If $\Gamma = \Psi_1$ (and $\Psi_0$, respectively), we check if Conditions (\romannumeral 1), (\romannumeral 2) and (\romannumeral 3) of the Global Convergence Theorem hold. First, Condition (\romannumeral 1) follows immediately from Condition 1.

For Condition (\romannumeral 2), we need to show that $g(\psi, Y_o)$ is a continuous function of $\psi$. In fact, for any sequence $\psi^{(t)} \rightarrow \bar\psi = (\bar M, \bar\psi)$, there exists an integer $T$ such that $M^{(t)} = \bar M$ for any $t>T$ since the model space $\mathcal M$ is finite. Hence, Condition 2 (\romannumeral 1) implies $g(\bar M, \theta^{(t)}, Y_o)\rightarrow g(\bar M, \bar\theta^{(t)}, Y_o)$. Furthermore, (a) and (b) follow from Assumption (\ref{des}) in this theorem and Theorem \ref{non-increasing}, respectively.

For $\psi^{(t)}_0 \rightarrow \psi_0$ and $\psi^{(t)}_1 \rightarrow \psi_1$ with $\psi^{(t)}_0 \in \mathbf A(\psi^{(t)}_1)$,  we have, by Condition 2 (\romannumeral 2), that
\begin{eqnarray}
Q(\psi^{(t)}_1; \psi^{(t)}_0) &= & Q(\psi^{(t)}_1; \psi^{(t)}_0) - Q(\psi_1; \psi^{(t)}_0) + Q(\psi_1; \psi^{(t)}_0) \rightarrow Q(\psi_1; \psi_0). \nonumber
\end{eqnarray}
Furthermore, we have $Q(\psi^{(t)}_0; \psi^{(t)}_0)\rightarrow Q(\psi_0; \psi_0)$ if we set $\psi^{(t)}_1 = \psi^{(t)}_0$. Since $Q(\psi^{(t)}_1; \psi^{(t)}_0) \leq Q(\psi^{(t)}_0; \psi^{(t)}_0)$, we have $Q(\psi_1; \psi_0) \leq Q(\psi_0; \psi_0)$; i.e., $\psi_1 \in \mathbf A(\psi_0)$. Thus, we have Condition (\romannumeral 3). $\hfill{} \Box$

{\it Proof of Corollary \ref{emsC1}}.
We set $\Gamma = \Psi_1$. For $\psi\not\in\Psi_1$ and $\tilde\psi\in \mathbf A(\psi)$ of EMS, we have $Q(\tilde\psi; \psi) < Q(\psi; \psi)$. Furthermore, we have $g(\tilde\psi, Y_o) < g(\psi, Y_o)$ by Theorem \ref{non-increasing}. Therefore, Assumption (\ref{des}) in Theorem \ref{gemsC} holds for the EMS algorithm. $\hfill{} \Box$

\section*{Appendix 2. Computing Q function in Gaussian graphical model}

Let $X_{ij}$ be the observation of th $i$ sample and th $j$ variable where $i = 1, \dots, n$ and $j = 1, \dots, p$.  Let $R = (R_{ij})\in \mathbb{R}^{n\times p}$ be an indictor matrix, where $R_{ij} = 0$ if $X_{ij}$ is missing, otherwise $R_{ij} = 1$.

To get the $Q$ function in the E-step, we first give the following notation and useful results. For $i = 1, \dots, n$, and $j, k=1,\dots, p$, we have
\begin{eqnarray}
E_t(X_{ij}) = \left\{\begin{array}{cc}
                             X_{ij} & \mbox{if } R_{ij}=1 \\
                             c_{ij} & \mbox{if } R_{ij}=0
                           \end{array}\right.\label{ey}\nonumber
\end{eqnarray}
and
\begin{eqnarray}
E_t(X_{ij}X_{ik}) = \left\{\begin{array}{ccl}
   X_{ij}X_{ik} & \mbox{if }& R_{ij}=R_{ik}=1 \\
   X_{ij}c_{ik} & \mbox{if }& R_{ij}=1, R_{ik}=0 \\
   \sigma_{i, jk} + c_{ij}c_{ik} & \mbox{if }& R_{ij}=R_{ik}=0
\end{array}\right..\nonumber
\end{eqnarray}
Here, $c_{ij} = (c_i)_j$, $c_{ik} = (c_i)_k$ and $c_i = \mu_{m_i} - \Omega_{m_im_i}^{-1}\Omega_{m_i o_i}(X_{i,o_i}-\mu_{o_i})$, and $\sigma_{i, jk} = ((\Omega^{-1})_{m_im_i})_{jk}$, $m_i$ and $o_i$ are the missing component and the observed component of the $i$ th observation, respectively. Let $E_t(\bar X) = n^{-1} E_t(X_i)$  and $S^* = n^{-1} \sum_{i=1}^n (X_i - E_t(\bar X))(X_i - E_t(\bar X))^\mathsf{T}$. The conditional expectation of $S^*_{jk}$ can be easily shown to be
\begin{eqnarray}
E_t(S^*_{jk}) &=& n^{-1}\sum_{i=1}^{n} E_t((X_{ij} - E_t(\bar X_j))(X_{ik} - E_t(\bar X_k)))\} \nonumber\\
&=&n^{-1}\sum_{i=1}^{n} E_t(X_{ij}X_{ik}) - E_t(\bar X_j) E_t(\bar X_k).\label{es}
\end{eqnarray}

For the Gaussian distribution $N(\mu_G, \Omega_G^{-1})$, the minus twice the log-likelihood is given by
\begin{eqnarray}
2 l(G, \mu_G, \Omega_G, X) = n\ln\det\Omega_G + \sum_{i=1}^n(X_i - \mu_G)^\mathsf{T}\Omega_G(X_i - \mu_G).\nonumber
\end{eqnarray}
The conditional expectation of $2 l(G, \mu_G, \Omega_G, X)$ is
\begin{eqnarray}
E_t (2 l(G, \mu_G, \Omega_G, X)) &=& n\ln\det\Omega_G + n\mathrm{tr}\big[\Omega_G E_t(S^*)\big]\nonumber \\
&& + n\mathrm{tr}\big[\Omega_G^{-1} (E_t(\bar X) - \mu_G)(E_t(\bar X) - \mu_G)^\mathsf{T}\big].\nonumber 
\end{eqnarray}
Thus, the $Q$ function can be obtained in (\ref{Q}).

{}

\end{document}